\documentclass[11pt]{article}
\pdfoutput=1
\usepackage{jheppub}
\usepackage{amsmath}
\usepackage{bm}
\usepackage{slashed}
\usepackage{graphicx}

\newlength{\dummysp}
\newcommand{\beq}{\begin{eqnarray}}
\newcommand{\eeq}{\end{eqnarray}}

\newcommand{\gappeq}{\mathrel{\rlap {\raise.5ex\hbox{$>$}}
{\lower.5ex\hbox{$\sim$}}}}
\newcommand{\lappeq}{\mathrel{\rlap{\raise.5ex\hbox{$<$}}
{\lower.5ex\hbox{$\sim$}}}}

\newcommand{\ben}{\begin{enumerate}}
\newcommand{\een}{\end{enumerate}}

\newcommand{\bit}{\begin{itemize}}
\newcommand{\eit}{\end{itemize}}

 \newcommand{\be}{\begin{equation}}
 \newcommand{\ee}{\end{equation}}
 \renewcommand{\Re}{\text{Re }}
  \renewcommand{\Im}{\text{Im }}
\def\[{\left [}
\def\]{\right ]}
\def\({\left (}
\def\){\right )}
\def\R{{\mathbb R}}
\def\S{{\mathbb S}}

\title{ 
{The absence of IR renormalons in gauge theories on $\mathbb R^3\times \S^1$ and what it means for resurgence}}
   
\author[1,2]{Mohamed M. Anber,} \author[3,4]{Tin Sulejmanpasic} 

\affiliation[1]{Institut de Th\'{e}orie des Ph\'{e}nom\`{e}nes Physiques, 
\'{E}cole Polytechnique F\'{e}d\'{e}rale de Lausanne, \\ CH-1015 Lausanne, 
Switzerland}
\affiliation[2]{Department of Physics, University of Toronto, 
Toronto, ON M5S 1A7, Canada}
\affiliation[3]{Department of Physics, North Carolina State University, Raleigh, NC, 27695, USA}
\affiliation[4]{Institute for Theoretical Physics, University of Regensburg, 93040 Regensburg, Germany}
\emailAdd{manber@physics.utoronto.ca}\emailAdd{tin.sulejmanpasic@physik.uni-regensburg.de}  
    
\abstract{We analyze the renormalon diagram of gauge theories on $\R^3\times \S^1$. In particular, we perform exact one loop calculations for the vacuum polarization in QCD with adjoint matter and observe that all infrared logarithms, as functions of the external momentum, cancel between the vacuum part and finite volume part, which eliminates the IR renormalon problem. We argue that the singularities in the Borel plane, arising from the topological neutral bions, are not associated with renormalons, but with the proliferation of the Feynman diagrams. As a byproduct, we obtain, for the first time, an exact one-loop result of the vacuum polarization which can be adapted to the case of thermal compactification of QCD.}

\begin{document}

\maketitle

\flushbottom
\section{Introduction}

Non-abelian gauge theories on $\R^3\times \S^1$ with center symmetry have been of great interest\cite{Unsal:2007jx,Poppitz:2012sw,Poppitz:2012nz,Poppitz:2013zqa,Misumi:2014raa,Anber:2014lba,Misumi:2014bsa} in the past years, not only as a tool for understanding QCD-like theories in a controlled, semi-classical regime, but potentially as a way to define a theory on $\R^4$ by arguing continuity in the compact radius $L$ of the circle \cite{Unsal:2008ch,Shifman:2008ja}. There has been tremendous progress in understanding the dynamics of the theory for $L\ll \Lambda{\mbox{\scriptsize QCD}}^{-1}$,  which is carried either by instanton-monopoles or bound states of instanton-monopoles known as \emph{bions}.

\begin{figure}[htbp] 
   \centering
   \includegraphics[width=0.5\textwidth]{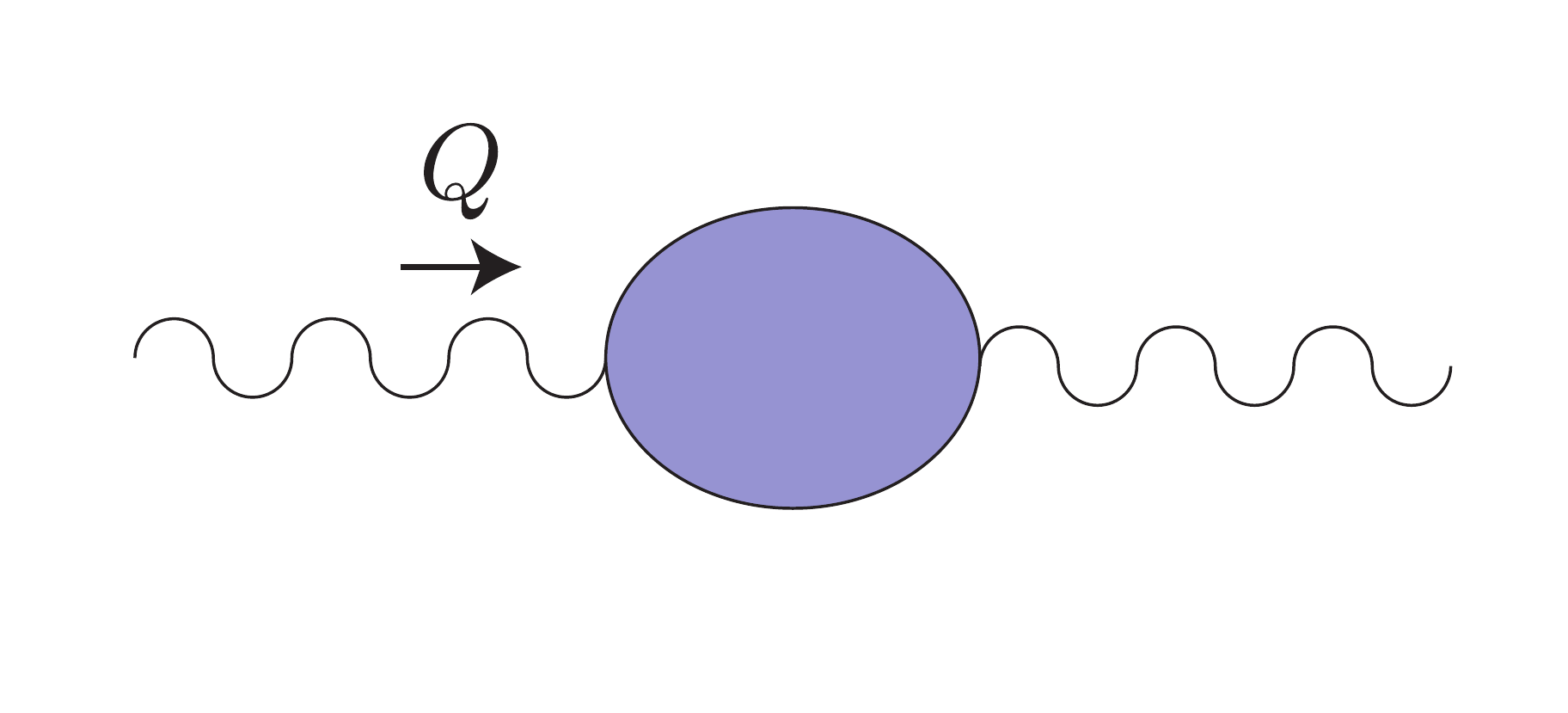}\includegraphics[width=0.5\textwidth]{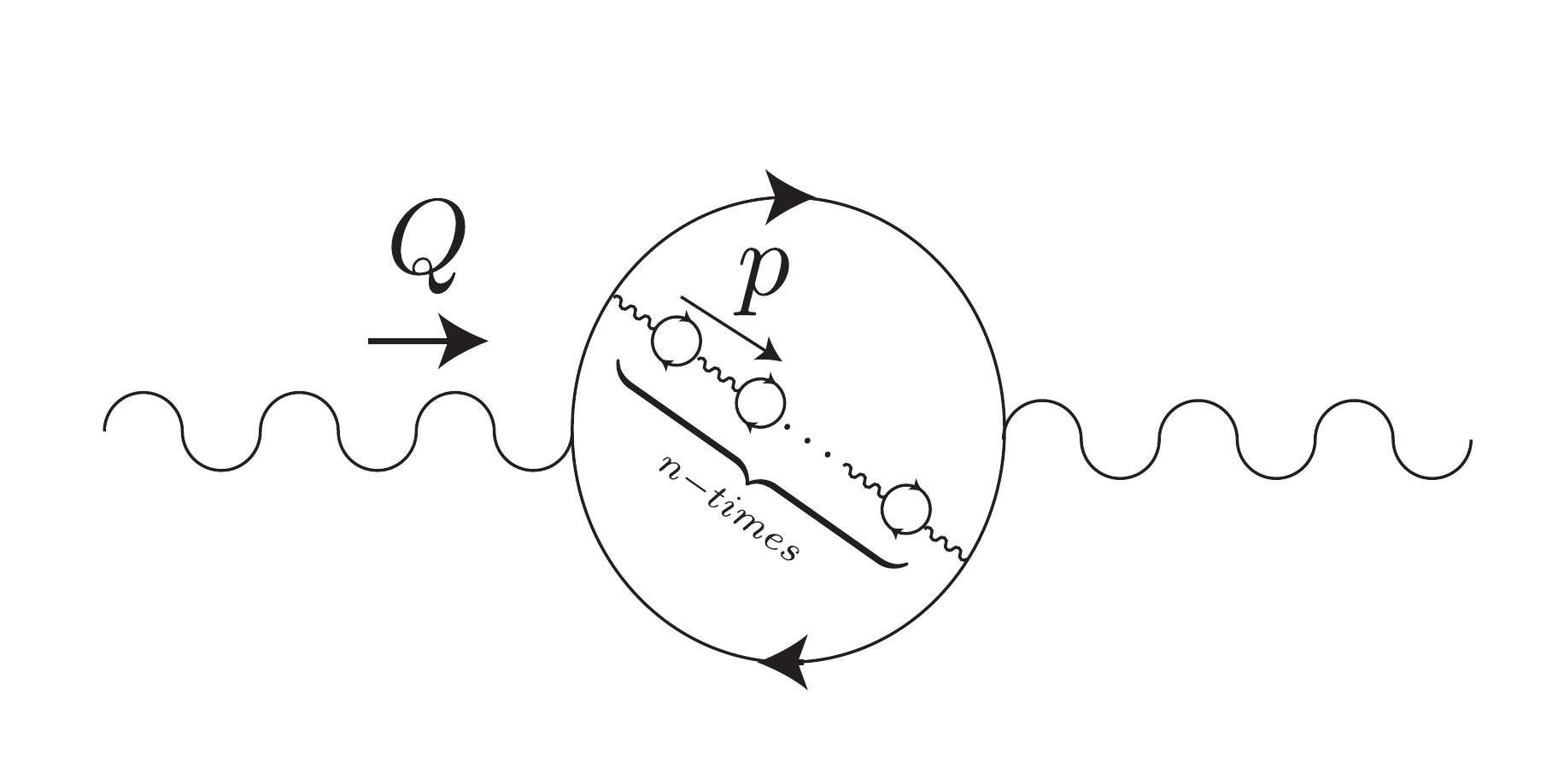} 
   \caption{Left: the vacuum polarization with all corrections. Right: Particular contribution to the vacuum polarization often referred to as the \emph{renormalon diagram}. }
   \label{fig:photon_full}
\end{figure}

On the other hand, non-abelian gauge theories on $\R^4$ are strongly coupled, and non-perturbative effects are notable. Even so, one may still hope that certain processes at short distance scales, or large momentum transfer\footnote{Capital letters are used to denote the 4-momenta,  and small letters denote the spatial 3-momenta.} $Q^2>> \Lambda_{\mbox{\scriptsize QCD}}^2$, are computable in perturbation theory. In a certain class of $n$ loop diagrams, however, the characteristic momentum running through the loops is not $Q^2$, but is exponentially suppressed with the number of loops\footnote{This suppression is caused by the appearance of logarithms in the one loop vacuum polarization diagrams (see Section \ref{IR renormalons: the sickness and cure} and \cite{Beneke:1998ui,Shifman:2013uka} for more details.), which we revisit in this work on $\R^3\times S^1$.} $n$. This suppression leads to $n!$ growth of the diagram upon integration over the momentum $P$ running through the chain of loops (see the right panel of Figure \ref{fig:photon_full}), rendering the loop expansion non-Borel summable (for review see \cite{Beneke:1998ui,Shifman:2013uka}). Another way of saying this is that the Borel plane contains poles on the real axis, which generate ambiguities in the calculation, depending on whether the pole is circumvented from above or from below.  The class of diagrams suffering from this problem are referred to as \emph{the renormalon diagrams} and the corresponding non-Borel summability is the (in)famous \emph{renormalon problem} \cite{'tHooft:1977am}. 

Borel non-summability of the perturbation theory is not in itself surprising and was argued by Dyson long time ago\footnote{Although it is true that the perturbation series is divergent, it was pointed out that Dyson's argument may not be entirely valid \cite{Bender:2001jr}.} \cite{Dyson:1952tj}. This problem also appears in quantum mechanics, but there the divergence is caused by the factorial proliferation of the number of the Feynman diagrams. In fact, one finds that such divergence is cured by instanton--anti-instanton events \cite{Bogomolny:1980ur,ZinnJustin:1981dx}, and has a priori  nothing to do with the renormalon problem. 

It was recently suggested in \cite{Argyres:2012vv,Argyres:2012ka} that IR renormalon ambiguity cancellation can be understood in terms of semi-classical instanton-monopole solutions appearing in the theory on $\R^3\times \S^1$, but which do not appear on $\R^4$. This idea was substantiated by the detailed analysis of two-dimensional models  on $\R\times \S^1$ \cite{Dunne:2012ae,Dunne:2012zk,Cherman:2013yfa,Cherman:2014ofa}, which have extra non-perturbative saddles compared to the theory on $\R^2$ (these are analogous to the instanton-monopoles in gauge theories). Since these theories reduce  to quantum mechanics for small $L$, a resurgent expansion can be constructed, in which case these saddles play a crucial role in canceling the ambiguities of the perturbation theory. It was conjectured in these works that these saddles (or rather their correlated pairs) cancel the  renormalon ambiguity which also exists in these theories on $\R^2$. This conjecture was made plausible because the leading renormalon ambiguity  matched the even-integer multiple of the action of the saddles, because the  beta-function coefficient, $\beta_0$, is an integer. This, however, is not the case in gauge theories in four dimensions, the beta-function coefficient $\beta_0$ is no longer an integer. This non-matching in gauge theories has led to the conjecture that renormalon ambiguities are shifted on $\R^3\times \S^1$ and that they smoothly interpolate from a semi-classical regime at $L\ll \Lambda_{\mbox{\scriptsize QCD}}^{-1}$ to those in the decompactification limit $L\rightarrow \infty$ \cite{Argyres:2012ka}.

In this work, however, we show that the renormalon problem vanishes completely from theories formulated on $\R^3\times \S^1$. In particular, we will show that all logarithmic dependence of the vacuum polarization (which we calculate exactly for arbitrary external momentum) in the center symmetric background cancel out, and hence no renormalon problem appears in the theory. In order to keep the center symmetry, we will mainly focus on Yang-Mills theory with adjoint matter, where center symmetry is perturbatively stable  \cite{Unsal:2007jx}.


For the theory on small $L\ll \Lambda_{QCD}^{-1}$, however the IR dynamics of QCD(adj) is governed by the instanton-monopoles and their correlated pairs which are called \emph{bions}. The neutral monopole--anti-monopole pairs (the \emph{neutral bions}) generate ambiguities which should be canceled by the perturbation theory and which were conjectured to be the semi-classical realization of renormalons \cite{Argyres:2012vv,Argyres:2012ka,Dunne:2012ae,Cherman:2013yfa}. Since we show that no renormalons exist in this theory, the inevitable conclusion is that the singularity in the Borel plane due to neutral bions is associated with the proliferation of diagrams, rather than the renormalon\footnote{Although the authors of \cite{Dunne:2012ae,Cherman:2013yfa,Cherman:2014ofa} do not stress this point, the cancellation between non-perturbative saddles and perturbation theory is clearly of this kind.}. On the other hand the factorial divergence of diagrams on $\R^4$ is believed to be associated to the 4D instanton--anti-instanton pairs. This is consistent with the fact that no factorial growth of diagrams happens in the large $N$ limit \cite{Koplik:1977pf}, as instantons have an exponentially vanishing contribution in this limit. The arguments leading to this conclusion should fail however on $\R^3\times \S^1$, provided that the large $N$ limit is taken with $NL$ kept fixed\footnote{\label{fn:abelian limit}Since the relevant scale in the center symmetric vacuum is $NL$, a naive 't Hooft limit would restore the $\R^4$ results. This is a form of large $N$ volume independence. The large $N$ limit with $NL$ kept fixed is often referred to the \emph{abelian large $N$ limit}.}, as the ambiguities of the neutral bions (which survive this kind of large $N$ limit) need to be canceled by an appropriate $n!$ growth. Since no renormalons exist on $\R^3\times \S^1$ it is natural to assume that this factorial behavior will come from some additional diagrams which appear in the theory on $\R^3\times \S^1$. 

Although the perturbation theory does not suffer from the IR renormalons, this should by no means be taken as an indication that the perturbation theory is complete. One immediate indication of this is that computation of the renormalon processes are very sensitive to the radius $L$, which they should not be since the full theory is gapped and should not feel the size of the box, so no physics should be affected by $L\gg \Lambda_{\mbox{\scriptsize QCD}}^{-1}$. However a question does pose itself: if all non-perturbative effects are systematically taken into account, would this $L$ dependence cancel in the final result? Of course there is an immediate problem here because the theory for large $L$ is not semi-classical. In the small $L\ll \Lambda_{\mbox{\scriptsize QCD}}^{-1}$ regime, however, the theory is semi-classical and under theoretical control \cite{Unsal:2007jx}, but the physical observables are not $L$-independent, even though there is some indication that  continuity between small and large $L$ holds\footnote{This statement must be made with care for $QCD(adj)$, as the continuous chiral symmetry will be broken for large $L$, but restored for small $L$. The observables calculated on small $L$, however, can be thought of as analytic functions of the number of adjoint (Weyl) flavors $n_W$ and $L$. In this case one can make a statement that the \emph{small $L$ and $n_W>1$ theory on $\R^3\times \S^1$ is continuously connected to the $n_W=0$ theory on $\R^4$ by taking the limit $L\rightarrow \infty$ $n_W\rightarrow 0$.}} on $\R^4$ \cite{Shifman:2008ja,Unsal:2008ch}. The $L$ dependence kicks in at the scale $NL\sim \Lambda_{\mbox{\scriptsize QCD}}^{-1}$, so that there is no reason to expect that the $L$ dependence of the renormalon diagram is canceled by the non-perturbative effects. This is even more so the case because of the fact that the regime of small $L$ is characterized by the hierarchy of scales for which $L\ll \text{(monopole separation)}$, so that the monopole screening of the perturbative propagators happens at the momentum scales much lower then the scale $1/L$, the point at which momentum dependence is cut off in the renormalon diagram. As $L$ is increased, however, these to scales (i.e. the monopole diluteness and the IR cutoff scale $L$) become of the order $\Lambda_{QCD}^{-1}$. This happens when the coupling is already strong, and perturbation theory is not justified anyway, so low momenta in Feynman diagrams should be cut off by non-perturbative effects. If complemented by the non-perturbative monopoles, however, the perturbation theory will have a new IR cutoff scale which is now as important as  $L$, so appropriate non-perturbative corrections need to be introduced in the perturbative propagators. 

It is perhaps important to stress that although the theory on $\R^3\times \S^1$ does not have renormalon poles in the Borel plane because the factorial growth $n!$ is cut off by the presence of the IR scale $L$, for $L\gg\Lambda_{QCD}^{-1}$ there will be some factorial growth which still may have physical meaning. Indeed it is perturbation theory complaining about the fact that one is using it in the regime where it should not be used. But since the unphysical growth of the series is cut off at large $n$, and since formally no singularity in the Borel plane exists due to the renormalon diagram, it is not as straightforward to attach meaning to this growth and to connect it to non-perturbative saddles. It seems clear, however, that whatever physics can be extracted in the regime of large $L\gg\Lambda_{QCD}^{-1}$, it should remain the same as on $\R^4$, due to the presence of the mass gap of order $\Lambda_{QCD}$. As the the radius $L$ is reduced and as the threshold $\Lambda_{QCD}^{-1}$ is reached, the semi-classical instanton-monopoles and bions will be the source of mass-gap and condensates, but no renormalon growth will be observed at all. On the other hand there will be factorial growth of diagrams associated with these saddles. Whether there is a connection between the two, seemingly different and completely unrelated factorial growths remains an open and important question.

These arguments are heuristic, however, but they do give hope that the theory on large $L$ can indeed be studied for small $L$, where all effects can be systematically accounted for and the large $L$ (and $n_W\rightarrow 0$) limit taken.

The article is organized as follows. In Section \ref{IR renormalons: the sickness and cure} we review the renormalon problem and how it arises in gauge theories on $\R^4$. We also qualitatively discuss why it is expected that no renormalons appear on $\R^3\times \S^1$. In Section \ref{Strategy and the calculation method} we introduce our computational strategy and obtain the general structure of the vacuum polarization tensors \emph{exactly} using the background field method for arbitrary external momentum. In Section \ref{Calculating the polarization tensor on a compact circle} we carry out systematic and {\em exact} calculations of the vacuum polarization tensor to one loop. To the best of our knowledge, this is the first exact calculations of the vacuum polarization diagram on $\mathbb R^3 \times \S^1$ and the results are easily applied to the case of thermal QCD and QED. Finally, in Section \ref{The static limit, resummation and  absence of IR renormalons} we show explicitly that no renormalons exist on $\mathbb R^3 \times \S^1$. We conclude in Section \ref{sec:conclusion} Various appendices summarize miscellaneous sums and integrals used in the computations. In particular, in Appendix \ref{eq:integrals_deriv} we use a novel method to obtain the exact result of new untabulated integrals. 

\section{IR renormalons: the sickness and cure}
\label{IR renormalons: the sickness and cure}

Although there are excellent reviews of renormalons on $\R^4$ \cite{Beneke:1998ui,Shifman:2013uka},  we will review the renormalon problem in $\R^4$ in this section for completeness. We also argue why this problem disappears when we formulate our theory on $\mathbb R^3 \times \S^1$ with preserved center symmetry. In this section, our arguments will be heuristic, postponing a careful analysis until the next section.

\subsection{IR renormalons on $\R^4$}
\label{sec:renormalons}

IR renormalons appear in processes which depend on a hard momentum scale $Q^2$. One such process is the gluon vacuum polarization graph (see left panel of Figure \ref{fig:photon_full}). Due to the gauge invariance, the vacuum polarization has the following structure
\be
\Pi_{\mu\nu}(Q)=(\delta_{\mu\nu}Q^2-{Q_\mu Q_\nu})\Pi(Q^2)\;.
\ee
The renormalons, however, are often discussed in the context of the so-called \emph{Adler function}, defined by
\be
D(Q^2)=4\pi^2 Q^2\frac{d\Pi(Q^2)}{dQ^2}\,.
\ee
The renormalon diagram is depicted in the right panel of Figure \ref{fig:photon_full}, and has the following form (see e.g.  \cite{Beneke:1998ui,Shifman:2013uka})\footnote{Notice that we have introduced the renormalization scale $\mu$ which must be taken as $\mu\gg \Lambda$ in order to insure small coupling and the validity of the perturbative expansion. Physical results, however, should not depend on $\mu$.}
\be\label{eq:adler4D}
D=\sum_{n=0}^\infty \alpha_s(\mu) \int_0^\infty \frac{dP^2}{P^2}F(P^2/Q^2)\left[\beta_{0,F}\alpha_s(\mu)\log\left(\frac{P^2}{\mu^2}\right)\right]^n\,,
\ee
where $\alpha_s(\mu)=\frac{g^2(\mu)}{4\pi}$ is the coupling at scale $\mu$, $P$ is the momentum which runs into the loop chain (see the right panel of Figure \ref{fig:photon_full}), and $\beta_{0,F}$ is the fermion contribution to the 1-loop $\beta$-function coefficient of the theory.
The exact expression for $F(P^2)$ can be found in \cite{Neubert:1994vb}, but its exact form is largely unimportant for the discussion of renormalons. What is important, however, is that for small $P^2/Q^2$ the function $F(P^2/Q^2)$ behaves as
\be
F(P^2/Q^2)\sim C P^{2a}/Q^{2a}+\dots
\ee
where $a=2$. Doing an integral in $P^2$ from $0$ to $Q^2$ gives the behavior
\be
D(Q^2)\sim \alpha_s(\mu)\sum_{n}\left(-\alpha_s(\mu)\frac{\beta_{0f}}{a}\right)^n n! \equiv S\,.
\ee
The Borel transform\footnote{The Borel transform of a series $S=\sum_{n}c_n$ is defined as $\mathcal B[S]=\sum_n c_n u^n/n!$. The original series can be recovered by $\int_0^\infty du\;e^{-u}\mathcal B[S]$.} of the above sum is
\be
{\cal B}[S]=\alpha_s(\mu)\sum_n \left(-\alpha_s(\mu)\frac{\beta_{0f}}{a}u\right)^n=\frac{\alpha_s(\mu)}{1+\alpha_s(\mu)\beta_{0f} u/a}\,,
\ee
and has a pole at $u=-\frac{a\beta_{0f}}{\alpha_s(\mu)}$. So far we have only considered fermion loops (hence the appearance of the $\beta_{0f}$). A convenient (and somewhat ad hoc) replacement of $\beta_{0f}\rightarrow \beta_{0}$ is often invoked, where $\beta_0$ is the full beta function coefficient of the theory, which, in asymptotically free theory, is negative. Hence the pole lies on the real axis, which in turn renders the sum non-Borel-summable. The pole can be circumvented from above or from below, which yields an imaginary ambiguity in the sum
\be
S=(\text{real part})\pm i\pi \frac{a}{\beta_0}e^{\frac{a}{\beta_0\alpha_s(\mu)}}\,.
\ee
Notice that the ambiguity is exponentially small and non-perturbative in the small coupling $\alpha_s(\mu)$. Using the one loop $\beta$-function and taking care of the prefactors, it is possible to show that
\be
\Im D(Q)\sim \left(\frac{\Lambda}{Q}\right)^4\;.
\ee
The above result offers an interpretation that there are certain condensates of order $\sim \Lambda^4$ beyond the perturbative treatment and that these condensates have an ambiguity which exactly cancels the renormalon ambiguity of the perturbative sector.

There is an alternative interpretation of the renormalon, however, which does not call for the introduction of the ambiguity advocated in \cite{Novikov:1984rf}. In this view the integration over momenta $p\lesssim \Lambda_{\mbox{\scriptsize QCD}}$ is nonsensical, as the propagators in this momentum regime would not be the simple, perturbative ones. In this more physical approach to the renormalon diagram, the momentum integral should be cut off in the infrared region at some scale $\mu$, sufficiently larger then $ \Lambda_{\mbox{\scriptsize QCD}}$ in order to render the perturbation theory valid. Since the scale $\mu$ is artificial and (almost) completely arbitrary, no physical observable should depend on it. On the other hand the condensates contain the momentum scales below the scale $\mu$, and therefore depend on $\mu$ as well. The consistent OPE should therefore involve the scale $\mu$ in the perturbative and in the condensate terms in such a way that $\mu$ cancels out completely from the final result.

Notice that this approach never encounters the factorial divergence of the loop summation, simply because all perturbative integrals are cutoff in the IR region of $p<\mu$, and its contribution is thrown into non-perturbative condensates. Therefore in this view the renormalon problem is gone and is viewed as the artifact of our lack of ability to probe the low momentum (i.e. strong coupling region) in perturbation theory. We will have little to say about this, admittedly more physical view, except that for large $L\gg \Lambda_{QCD}$ on $\R^3\times \S^1$ these arguments should not change. 

\subsection{Overview of the theory on $\R^3\times \S^1$ and the absence of the renormalon problem}\label{sec:overview}

Let us present some heuristic arguments of why no IR renormalons are expected in the center symmetric theory on $\R^3\times \S^1$. Our exposition, as well as the formulas,  will be very schematic and we postpone detailed calculation of vacuum polarization on $\R^3\times \S^1$ for the sections to follow. We give our argument for the $SU(2)$ case, but it applies trivially to higher rank gauge groups. 

The theory we discuss is described by the Lagrangian \eqref{eq:action} with  $n_W$ Weyl fermions in the adjoint representation. If center symmetry is preserved then the vacuum configuration that one needs to expand around is $A_3=\frac{v}{2}\tau^3$ (we chose the third spatial direction to be compact), with $v=\pi/L$, which is the center symmetric point stabilized perturbatively\footnote{Although in this work we use QCD(adj) to do all the calculations, the conclusions we give apply equally to any QCD-like theory with a stable center symmetry. In a generic theory, however, some non-perturbative terms need to be included in order to render the center stable, and the ``electric mass'' non-tachyonic.} in QCD(adj) and we have chosen a gauge in which $A_3$ is always in the third color direction.

In such a vacuum, the gauge symmetry is spontaneously broken to $U(1)$ and one can distinguish between gauge and fermion fields in the third color direction whose  propagators remain massless, and fields in the $1$ and $2$ color directions which get a mass of order $1/L$. The gauge field along the the third color direction is the $U(1)$ photon denoted by $a_{m}, m=0,1,2,3$, for which the longitudinal component gets a mass by one loop effects. For small $L\ll \Lambda_{\mbox{\scriptsize QCD}}^{-1}$ the theory is in a weakly coupled regime and hence  one can apply reliable semi-classical analysis to find that the theory is gapped due to the proliferation of non-perturbative and non-self-dual topological molecules known as magnetic bions \cite{Unsal:2007jx}. For large $L$, although no abelianization can be invoked, from the point of view of perturbation theory the propagators fall into a class of massless, would-be $U(1)$ photon, and massive would-be $W$-bosons, whose low momentum dependence in the propagator is cut off at $1/L$. In this sense the radius $L$ serves as an IR regulator for the $W$-boson propagators. 

First, we discuss the massive gauge bosons. Looking at the renormalon diagram in Figure \ref{fig:photon_full}, we consider the case when all wavy lines are massive gauge bosons either $W$-bosons or longitudinal $a_3$ (we denote the compact direction by the $3$-component). Then the propagators in the renormalon diagram are all well behaved in the IR and are of the form $\sim 1/(p^2+m^2)$, while the fermion loop has a structure $f(p)$ which vanishes\footnote{If the function $f(0)=const$, then one must first subtract the constant and absorb it into the mass $m^2$} when\footnote{Since we are interested in the low $p$-momentum behavior, we only analyze the case  $p_3=0$, i.e. the zero Matsubara mode. The higher Matsubara modes cannot cause problems in the IR as the Matsubara frequency acts as an IR regulator.} $p\rightarrow 0$. The renormalon diagram then has a structure
\be
\int d(p/q) \;F(p/q)\left(\frac{f(p)}{p^2+m^2}\right)^{n}
\ee
where $F(p)$ is some function of $p$ which is not pathological in the IR\footnote{Note that this is not the same function as in Section \ref{sec:renormalons}, and that we have not computed this function which comes from integrating over the momentum in the ``large fermion loop'' of the renormalon diagram. However no strange IR behavior should arise from this computation. In fact the calculation can be dramatically simplified if the assumption $qL\gg 1$ is made, in which case the Matsubara sum over the large fermion loop can be converted into an integral. However, since this function determines the position of the renormalon pole, its structure may still be important. See Section \ref{sec:conclusion}}. Since the expression $f(p)/(p^2+m^2)$ vanishes for small $p$, the $p$-integral is cutoff rapidly at low momenta for large $n$, and hence small $p$ contribution to the integral is negligible, and no factorial dependence arises from this integral. We will remind the reader that in the case of the analogous computation on $\R^4$ the situation was different as $m=0$ and $f(P)=P^2\log P^2$, and it was the appearance of the softly IR divergent $(\log P)^n$ term in the integrand which lead to the $n!$ behavior. Here however the integrand is made perfectly well behaved in the IR because of the mass term.

Next let us discuss the case of massless photon in the renormalon diagram. In this case the wavy lines in Figure \ref{fig:photon_full} are all massless photons. However, the fermions in the loop have to be charged under the $U(1)$ and are heavy with the mass of order $\sim 1/L$. Then the loop can be sensitive to the external momentum only up to the IR scale $1/L$. To say it differently, the coupling constant for the massless photon stops running perturbatively once the scale $L^{-1}$ is reached, as the only way this running can occur is through the mediation of the massive $W$-bosons and massive fermions which are not in the third color direction. This means that the vacuum polarization in the IR reduces to
\be
\Pi_{ij}=(\delta_{ij}p^2-p_i p_j)\times \text{(constant)}\;.
\ee 
which results in the following structure of the Adler function:
\be
\int d(p/q)\; F(p/q) \text{(constant)}^n\;.
\ee
Again no factorial behavior $n!$ is observed. Therefore renormalon problem, in the formulation which we gave in Sec. \ref{sec:renormalons} does not exist on $\R^3\times \S^1$. 

 Let us give another, more quantitative, argument of why IR logarithmic singularities of the vacuum polarization disappear on $\R^3\times\S^1$. The key observation is to note that the result of the Matsubara sums can be split into two parts: a vacuum contribution  and ``thermal'' excitations contribution\footnote{Of course, since fermions are periodic in the compact direction, no thermal interpretation holds for this setup.}.  The ``thermal'' part is characterized by an appearance of Bose-Einstein distribituion\footnote{This factor is really just the Bose-Einstein--like thermal excitation factor of particles who's Boltzmann factor is $e^{-kL}$ and coupling to the $A_3$ abelian $U(1)$ field is $e^{\pm i\int A_3dx^3}$, so they carry ``electric charge''.} $\Re\frac{1}{e^{kL+i\mu}-1}$, where $\mu=vL$ is the holonomy, while the vacuum part can be obtained by a replacement $\Re\frac{1}{e^{kL+i\mu}-1}\rightarrow 1/2$ (see e.g. \eqref{the first sum} and \eqref{the second sum}). The holonomy appears in these calculations because the particles running in the loop are charged under the remaining $U(1)$ gauge group. However, notice that $\lim_{k\rightarrow 0}\Re \frac{1}{e^{kL+i\mu}-1}=-\frac{1}{2}$, so that in the IR the vacuum and the ``thermal'' contribution cancel, and the loops are better behaved for low momenta then they are on $\R^4$. It is this observation which causes the IR logarithms to cancel between the thermal and the vacuum part (for details the reader is referred to the Section \ref{Calculating the polarization tensor on a compact circle}).

The argument given above is valid only for $\mu\ne 0\mod 2\pi$. For $\mu=0\mod 2\pi$ the divergence is actually worse than logarithmic, but logarithms still exist and they still still cancel\footnote{This cancellation was also demonstrated explicitly for $\mu=0\mod 2\pi$ in \cite{Gross:1980br}, and to which all our expressions reduce in this limit.}. To see this we computed the necessary integrals exactly which are valid for any $\mu$, even for $\mu=0\mod 2\pi$.

In the rest of the paper we make the picture portrayed above more precise by carrying out detailed  calculations for the vacuum polarization diagrams for the massless photon to one loop.

\section{Strategy and the calculation method}
\label{Strategy and the calculation method}

In this section, we explain the elements of the method we use to calculate the one-loop vacuum polarization. Since we perform our calculations for QCD(adj), we first summarize the perturbative dynamics of this theory in subsection \ref{Dynamics of QCD(adj) over a circle}. As was mentioned in the introduction, the renormalon calculations start by assuming a large number of fermion flavors running in the loops. In this case, the type of diagrams  depicted in the right panel of Figure (\ref{fig:photon_full}) will be enough to show the $n!$ growth associated with the appearance of renormalons. However, including the non-abelian contribution is far more complicated. For example, adding the gluon and the ghost bubbles to the fermion bubble is not sufficient to guarantee a gauge invariant answer. In fact, in order to respect the gauge invariance of the theory,  the number of diagrams we need to calculate proliferate making any attempt to perform such calculations impractical.  In order to circumvent this problem, we use a convenient computational device by replacing the one-loop running of the coupling on $\R^3\times \S^1$ due to fermions with the full running of the coupling. In turn, this reduces the hard problem to a simpler one: we just need to calculate the one-loop correction to the running of the coupling constant on $\mathbb R^3 \times \mathbb S^1$ in the presence of a non-trivial holonomy. In order to avoid considering vertex correction, a convenient way to perform such calculations is to use the background field method where only vacuum polarization diagram needs to be computed. In the following subsection, we explain this method adapted to our geometrical setup. After obtaining the one-loop vacuum polarzation, one then needs to sum a series of bubbles to obtain the full propagator. This summation process is reviewed in subsection \ref{The general form of the gluon propagator}. 

Throughout this work, we perform our analysis in the Euclidean space and we use the imaginary time formalism (the Matsubara technique)  to carry out our calculations. We also use elements of the Lie algebra for the sake of generality, but we focus mainly on the cases of $su(2)$ and $su(3)$ algebra where expressions simplify at the center symmetric point. Generalizing our results to a general gauge group is straightforward. We use capital letters $P,K$ to denote four-dimensional Euclidean quantities and boldface letters to denote their three-dimensional component $\pmb p, \pmb k$ such that $P=(p_0,\pmb p)$. The magnitude of the three dimensional quantities will be denoted by normal letters  $p\equiv|\pmb p|$. Notice also that we use boldface letters to denote quantities that live in the Cartan subspace. It will be obvious from the context which structure we mean.

\subsection{ Perturbative dynamics of QCD(adj) on $\mathbb R^3 \times \mathbb S^1$ }
\label{Dynamics of QCD(adj) over a circle}

We consider Yang-Mills theory on $\mathbb R^3 \times \mathbb S^1$ with a compact gauge group $G$ and $n_W$ massless Weyl fermions. We compactify the $x_3$ direction such that $x_3\sim x_3+L$, where $L$ is the circumference of the $\mathbb S_1$ circle. The  action of the theory is given by
\begin{eqnarray}\label{eq:action}
S=\int_{\mathbb R^3\times \mathbb S^1} \mbox{tr}\left[\frac{1}{2g^2}f_{mn}f^{mn}-2i\bar \lambda_I \bar \sigma^m D_m \lambda_I \right]\,,
\end{eqnarray} 
where $f_{mn}=\partial_m a_n-\partial_n a_m+i[a_m,a_n]$. is the field strength tensor  and $I$ is the flavor index. In this paper, the  letters, $m,n$ run over $0,1,2,3$, while the Greek letters $\mu,\nu$ run over $0,1,2$.  We also write $a_m\equiv a_m^a t^a$ and $\lambda\equiv \lambda_a t^a $, where $t^a$ are the Lie algebra generators and the letters $a,b$ denote the color index. A brief review of a few elements of Lie algebra used in this paper is provided in Appendix  \ref{Appendix propagators}. The $x_0$ axis is the time direction, and hence the compact direction $x_3$ is one of the spacial directions. Therefore, both gauge bosons and fermions obey periodic boundary conditions around $\mathbb S^1$.

The quantum theory has a dynamical strong scale $\Lambda_{\mbox{\scriptsize QCD}}$ such that to one-loop order we have
\begin{eqnarray}
g^2(\mu)=\frac{16\pi^2}{\beta_0}\frac{1}{\log\left(\mu^2/\Lambda_{\mbox{\scriptsize QCD}}^2\right)}\,,
\end{eqnarray}
where $\mu$ is a normalization scale, $\beta_0=\frac{(11-2n_W)c_2}{3}$, and $c_2$ is the dual Coexter number which is equal to $N$ for $su(N)$ algebra. In this work we will take $n_W<5.5$ so that our theory is asymptotically free. Moreover, by working at a small spacial circle, i.e. for $L\Lambda_{\mbox{\scriptsize QCD}}<<1$ we find that the coupling constant remains small and hence we can perform reliable perturbative calculations. Therefor, for $L\Lambda_{\mbox{\scriptsize QCD}}<<1$, we can use perturbation theory to integrate out the Kaluza-Klein tower of the gauge fields and fermions. This can be performed in a self-consistent way by first writing down the components of the gauge fields and fermions  in  the Weyl-Cartan basis (see Appendix  \ref{Appendix propagators}):
\begin{eqnarray}
X=X^at^a=\pmb X\cdot \pmb H+\sum_{\pmb \beta _+}X_{\pmb \beta}E_{\pmb \beta}+\sum_{\pmb \beta _+}X_{\pmb \beta}E_{-\pmb \beta}\,,
\end{eqnarray}
where $\pmb X=(X^1,X^2,...,X^r)$ denotes the Cartan components of any field, and $\{\pmb\beta_+\}$ is the set of positive roots. We use boldface letters to denote vectors in the Cartan subspace of the color space. Later in this work, we will also use boldface letters to denote three dimensional vectors in the Euclidean space. This should not bring on any confusion since it will be clear which space we mean.  Next, we assume that the quantum corrections will induce a vacuum expectation value for the gauge fields along the $x_3$ direction. Defining  $\pmb A_3\equiv \frac{\pmb\phi}{L}$, we find that for a general value of $\pmb\phi$, the gauge group $G$ spontaneously breaks down into its $U(1)^r$ subgroups, where $r$ is the rank of the group. In this case, the bosonic part of the dimensionally reduced action on $\mathbb R^3$ reads
\begin{eqnarray}
S=\int_{\mathbb R^3} d^3 x\left\{\frac{L}{2g_g^2 }\frac{\partial_\mu \pmb \phi\cdot \partial_\mu \pmb \phi}{L^2}+\frac{L}{4g_s^2}\pmb f_{mn}^2+V_{\mbox{\scriptsize eff}}(\pmb \phi) \right\}\,. 
\end{eqnarray}
This is the long distance effective action on the three dimensional space. Heavier fields are of order $1/L$ and their effects will show up as corrections to the classical action. 
In fact, the scalar field $\pmb \phi$ is the gauge field component along the $x_3$ direction, and  its effective potential $V_{\mbox{\scriptsize eff}}(\pmb \phi)$ results from quantum corrections to this field. On the other hand, an effective potential to the field $\pmb v_{\mu\nu}$ is forbidden thanks to the $U(1)^r$ gauge symmetry. However, quantum corrections will result in a wave function renormalization which in turn will  modify the value of the coupling constants $g_g$ and $g_s$, which in general are different.

The quadratic term in $V_{\mbox{\scriptsize eff}}(\pmb \phi)$ can be obtained by computing the two-point function as we will do in section \ref{The static limit, resummation and  absence of IR renormalons}. However, one can also obtain the full result of the potential by performing Gross-Pisarski-Yaffe on-loop analysis:
\begin{eqnarray}
V_{\mbox{\scriptsize eff}}(\pmb \phi)=(-1+n_W)\frac{4}{\pi^2 L^4}\sum_{n=1}^\infty \sum_{\pmb\beta_+}\frac{\cos(n\pmb \beta \cdot \pmb \phi)}{n^4}\,.
\end{eqnarray}
In this work, we will be interested in the $su(N)$ group, and specifically $N=2,3$ only. In this case the minimum of the potential $V_{\mbox{\scriptsize eff}}(\pmb \phi)$ is located at
\begin{eqnarray}
\pmb \phi_0=\frac{2\pi \pmb\rho}{N}\,,
\label{point of center symmetry}
\end{eqnarray}
where  $\pmb \rho$ is the Weyl vector $\pmb \rho=\sum_{u=1}^{r=N-1}\pmb\omega_u$, $\pmb\omega_u$ are the fundamental weights which satisfy $\pmb\omega_u\cdot \pmb \alpha_v=\delta_{uv}$, and  $\pmb \alpha_u$ are the simple roots. At these values of $\pmb\phi_0$, one can easily check that the $\mathbb Z_{N}$ center symmetry of the $SU(N)$ gauge group is preserved.

\subsection{The background field method on $\mathbb R^3 \times \mathbb S^1$ in the presence of non-trivial holonomy}

\begin{figure}[b] 
   \centering
   \includegraphics[width=3in]{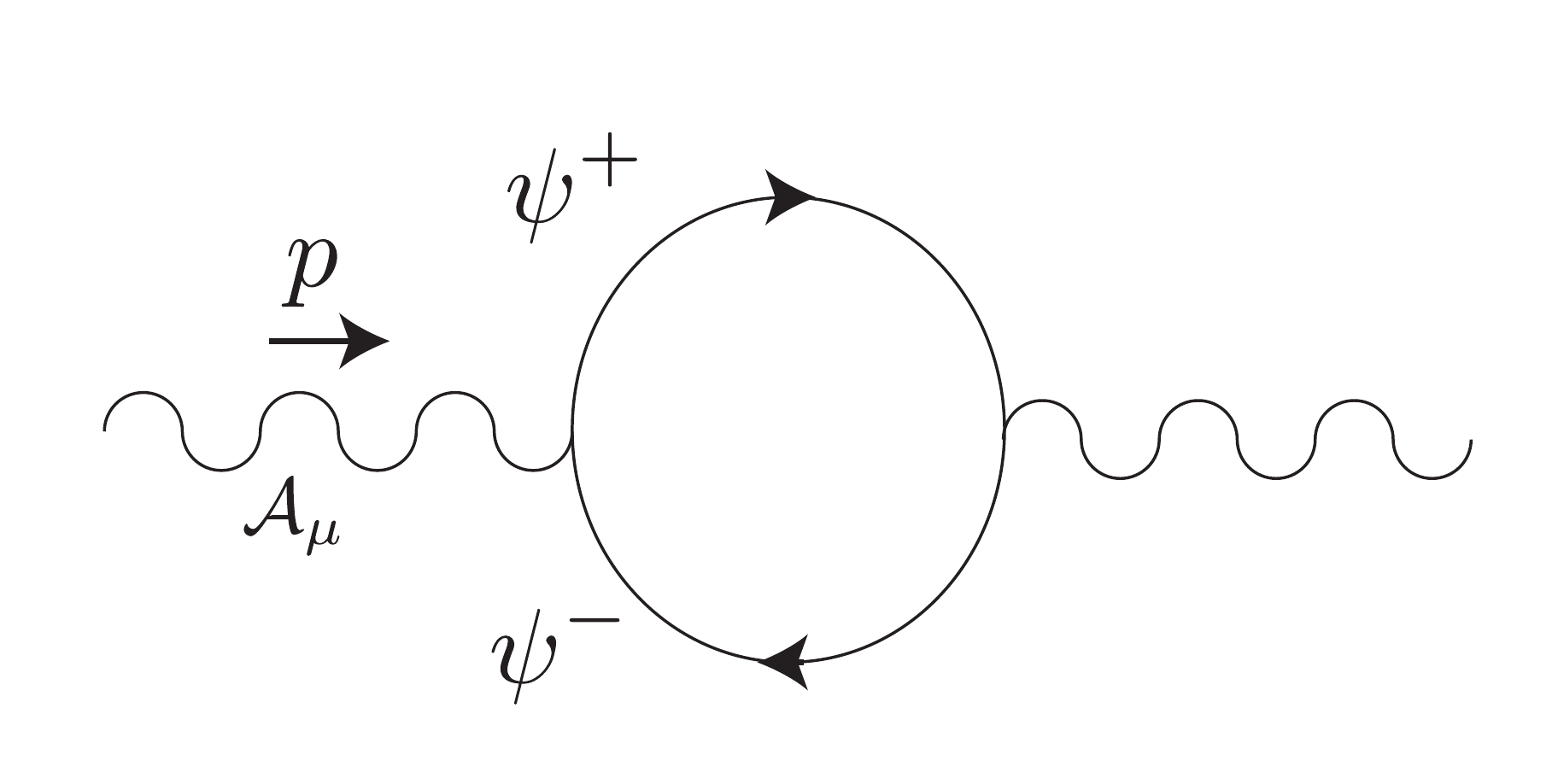} 
   \caption{Fermion contribution to the vacuum polarization.}
   \label{fig:vac_polarization_ferm}
\end{figure}

The background field method is a powerful tool to compute the quantum corrections without losing the explicit gauge invariance of the theory. The essentials of this method goes back to the sixties of the last century DeWitt \cite{DeWitt:1967ub}. In this method one writes the gauge field appearing in the classical Lagrangian as ${\cal A}+a$, where ${\cal A}$ is the classical background field, and $a$ is the quantum fluctuations. In 1980, Abbott \cite{Abbott:1980hw} showed how to generalize the background field method to include multi-loops, and he gave explicit prescription including Feynman rules to compute gauge-invariant effective action. In this work, we use the same technique and rules as given by Abbott in order to calculate the one-loop correction to the gluon propagator in the presence of non-trivial holonomy for any gauge group $G$. We explain in details how to do this for adjoint fermions, and then we describe a simple recipe to include the contributions from the non-abelian gauge fields.   Throughout this section, we need to use elements of the Lie algebra technology. In addition, we need the expressions of the propagators on $\mathbb R^3 \times \mathbb S^1 $ in the presence of a non-trivial holonomy. Both of these topics are summarized in Appendix \ref{Appendix propagators}.

The adjoint Dirac fermions couple to the holonomy $A_3$, the dynamical gluon $a_m$, and the background field ${\cal A}_m$ as
\begin{eqnarray}
{\cal L}=\int d^4 x\bar \psi^a\gamma_m \left(\partial_m\delta^{ac}+f^{abc}A^b_0\delta_{0m}+f^{abc}a^b_m+f^{abc}{\cal A}^{b}_m\right)\psi^c\,,
\end{eqnarray}
from which one can easily read the fermion-background field vertex:
\begin{eqnarray}
gf^{abc}=-ig(T_{\mbox{\scriptsize adj}}^b)_{ac}\,.
\end{eqnarray}
Then, using the propagator (\ref{fermion prop first form}), the one-loop fermion contribution to the vacuum polarization for any gauge group reads (see Figure \ref{fig:vac_polarization_ferm}) 
\begin{eqnarray}
\nonumber
&&\Pi^{\mbox{\scriptsize D}\,ed}_{mn}(p, \omega)=\\
\nonumber
&-&\frac{g^2}{L}\sum_{q\in \mathbb Z}\int\frac{d^3 k}{(2\pi)^3}\mbox{Tr}_{\mbox{\scriptsize adj}}\left[T^eT^d\gamma_m \frac{1}{\slashed{K}}\gamma_{n}\frac{1}{\slashed{K}+\slashed{P}} \right]\\
\nonumber
&=&-\frac{4g^2}{L}\sum_{q\in \mathbb Z}\mbox{tr}_{\mbox{\scriptsize adj}}\left\{\int\frac{d^3 k}{(2\pi)^3}\frac{T^eT^d\left[\delta_{mn}\left(-K\cdot P-K^2\right)+K_m P_n +K_n P_m+2K_m K_n\right]}{\left[k^2+\left(\frac{2\pi q}{L}+A_3^bT^b\right)^2 \right]\left[(\pmb k+\pmb p)^2+\left(\frac{2\pi q}{L}+A_3^bT^b+\omega\right)^2 \right]}\right\}\,.\\
\label{Pi for fermions all groups}
\end{eqnarray}
where the superscript $D$ denotes that the expression is given for a single Dirac fermion, Tr denotes the trace over both the color and gamma matrices, $K\cdot P=k_0\cdot p_0+\pmb  k\cdot \pmb p$, $K^2=k_0^2+k^2$, $k_0=\frac{2\pi n}{L}+A_3^bT^b$, and $p_0=\omega$.

One can also use the background field method in the Cartan-Weyl basis by turning on background fields along the Cartan generators ${\cal A}_m={{\cal A}}_m^i  H^i$. In this case, one can derive the  fermion-background field vertex simply by replacing $\beta^i  \phi^i $ with $\beta^i{{\cal A}}_\mu^i$ in (\ref{final expression for fermions Lagrangian}). Thus, we find the fermion-background field vertex:
\begin{eqnarray}
g\beta^i {{\cal A}}^i_\mu\,.
\end{eqnarray}
Then, we use the propagator (\ref{fermion prop second form}) to find the vacuum polarization in the Cartan-Weyl basis:
\begin{eqnarray}
\nonumber
\Pi^{\mbox{\scriptsize D}\,ij}_{mn}=-\frac{4g^2}{L}\sum_{\pmb \beta}\sum_{q\in \mathbb Z}\int\frac{d^3 k}{(2\pi)^3}\frac{\beta^i \beta^j\left[g_{mn}\left(-K\cdot P-K^2\right)+K_m K_n +K_n P_m+2K_m K_n\right]}{\left[k^2+\left(\frac{2\pi q}{L}+\frac{\phi^i\beta^i}{2\pi L}\right)^2 \right]\left[(\pmb k+\pmb p)^2+\left(\frac{2\pi q}{L}+\frac{\phi^i\beta^i}{2\pi L}+\omega\right)^2 \right]}\,.\\
\label{Pi fermion in Cartan}
\end{eqnarray}

In order to simplify the calculations,  we take $G=su(N)$ and perform our analysis at center-symmetry. The value of the holonomy in the center-symmetric case is given by \eqref{point of center symmetry}. In addition,  for $N=2,3$ one can easily see that the combination $\frac{2\pi n+\rho^i  \beta^i}{L}$ falls into one of two categories: either $\frac{2\pi n+\mu}{L}$ or $\frac{2\pi n-\mu}{L}$, 
\footnote{This is trivial to see in $su(2)$ since there are only two roots $\pmb \beta_{1,2}=\pm1$. The $su(3)$ case requires a bit more work. The roots, fundamental  weights, and Weyl vector are given by
\begin{eqnarray}
\nonumber
\pmb\beta_{1}&=&\left(1,0\right)\,,\pmb\beta_{2}=\left(-\frac{1}{2},\frac{\sqrt 3}{2}\right),\,\pmb\beta_{3}=\left(-\frac{1}{2},-\frac{\sqrt 3}{2}\right)\,,\\
\nonumber
\pmb\beta_{4}&=&\left(-1,0\right)\,,\pmb\beta_{5}=\left(\frac{1}{2},-\frac{\sqrt 3}{2}\right),\,\pmb\beta_{6}=\left(\frac{1}{2},\frac{\sqrt 3}{2}\right)\,,\\
\pmb\omega_1&=&\left(1,\frac{1}{\sqrt 3}\right)\,,\pmb\omega_2=\left(0,\frac{2}{\sqrt 3}\right)\,,\pmb \rho=\left(1,\sqrt 3 \right)\,.
\end{eqnarray}
Then, we find $2\pi\pmb \rho \cdot \pmb \beta_{1}/N=2\pi/3$, $2\pi\pmb \rho \cdot \pmb \beta_{2}/N=2\pi/3$, and $2\pi\pmb \rho \cdot \pmb \beta_{3}/N=-4\pi/3$. We see that theses values belong to the first category given that we shift $n\rightarrow n+1$ in $\frac{2\pi n+\pmb\rho\cdot \pmb \beta_3}{L}$. Similarly we find that the values  $2\pi\pmb \rho \cdot \pmb \beta_{4}/N=-2\pi/3$, $2\pi\pmb \rho \cdot \pmb \beta_{5}/N=-2\pi/3$, and $2\pi\pmb \rho \cdot \pmb \beta_{6}/N=4\pi/3$ belong to the second category after making the shift  $n\rightarrow n-1$ in $\frac{2\pi n+\pmb\rho\cdot \pmb \beta_6}{L}$.} 
where 
\begin{eqnarray}
\mu=\frac{2\pi}{N}\,,\quad N=2,3.
\end{eqnarray}
Therefore, for $su(2)$ and $su(3)$ the vacuum polarization tensor takes the form
\begin{eqnarray}
\nonumber
\Pi^{\mbox{\scriptsize D}\,ij}_{mn}&=&-\frac{4g^2}{L}\sum_{\pmb \beta_{(1)}}\sum_{q\in \mathbb Z}\int\frac{d^3 k}{(2\pi)^3}\frac{\beta^i\beta^j\left[\delta_{mn}\left(-K\cdot P-K^2\right)+K_m P_n +K_n P_m+2K_m K_n\right]}{\left[k^2+\left(\frac{2\pi q+\mu}{L}\right)^2 \right]\left[(\pmb k+ \pmb p)^2+\left(\frac{2\pi q+\mu}{L}+\omega\right)^2 \right]}\\
\nonumber
&-&\frac{4g^2}{L}\sum_{\pmb \beta_{(2)}}\sum_{q\in \mathbb Z}\int\frac{d^3 k}{(2\pi)^3}\frac{\beta^i\beta^j\left[\delta_{mn}\left(-K\cdot P-K^2\right)+K_m P_n +K_n P_m+2K_m K_n\right]}{\left[k^2+\left(\frac{2\pi q-\mu}{L}\right)^2 \right]\left[(\pmb k+\pmb p)^2+\left(\frac{2\pi q-\mu}{L}+\omega\right)^2 \right]}\,,
\end{eqnarray}
where $\pmb \beta_{(1,2)}$ denotes the roots in the first or second category. Now using $\sum_{\pmb\beta_{(1,2)}}\beta^i\beta^j=\delta^{\mu\nu}N/2$ (keeping in mind that $N=2,3$ only), we finally obtain 
\begin{eqnarray}
\nonumber
\Pi^{\mbox{\scriptsize D}\,ij}_{mn}(p, \omega)&=&-\delta^{\mu\nu}\frac{4Ng^2}{2L}\sum_{q\in \mathbb Z}\int\frac{d^3 k}{(2\pi)^3}\frac{\delta_{mn}\left(-K\cdot P-K^2\right)+K_m P_n +K_n P_m+2K_m K_n}{\left[k^2+\left(\frac{2\pi q+\mu}{L}\right)^2 \right]\left[(\pmb k+\pmb p)^2+\left(\frac{2\pi q+\mu}{L}+\omega\right)^2 \right]}\\
&+&(\mu \rightarrow -\mu).
\label{final expression Pi in cartan}
\end{eqnarray}
It is easy to understand the physics behind the simplified formula (\ref{final expression Pi in cartan}) for $N=2,3$ since in these cases all charged fermions $\psi_{\pmb \beta}$ have exactly the same mass in the center-symmetric vacuum. For $N>3$ the charged fermions will generally have different masses, and hence one has to restore to the original expressions (\ref{Pi for fermions all groups}) or (\ref{Pi fermion in Cartan}).
 
\begin{figure}[t] 
   \centering
   \includegraphics[width=.85\textwidth]{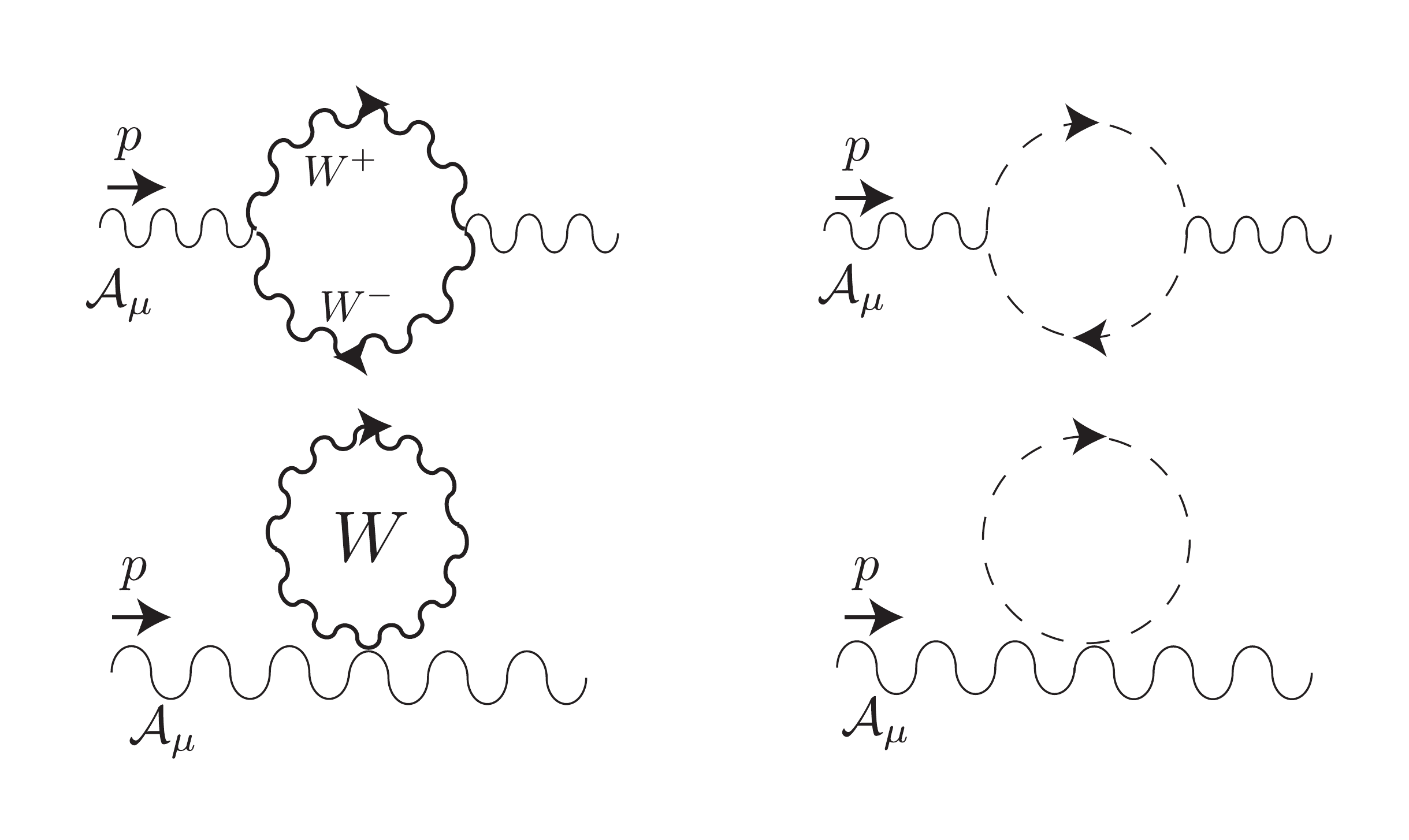} 
   \caption{Diagrams contributing to the non-ableian part of the vacuum polarization \eqref{the nonabelian polarization}.  The dashed lines are the ghosts, and that the second line of diagrams does not contribute on $\R^4$ in a normalization that doesn't violate gauge invariance.}
   \label{fig:vac_pol_gauge_sector}
\end{figure}
 
In fact, one can obtain the result (\ref{final expression Pi in cartan}) directly from (\ref{Pi for fermions all groups}) for $N=2,3$ by setting $A_3^bT^b=\pm \mu/L$ and using $\mbox{Tr}_{\mbox{\scriptsize{adj}} }[T^eT^d]=f^{adc}f^{cea}=N\delta^{ed}$. This is a huge simplification since one can then use the same background field Feynman rules, as given by Abbott \cite{Abbott:1980hw}, to compute the non-abelian one-loop corrections on $\mathbb R^3 \times \mathbb S^1$, shown in Figure \ref{fig:vac_pol_gauge_sector}, provided that we substitute the ghosts and gluons propagators on $\mathbb R^4$ with the propagators  on $\mathbb R^3 \times \mathbb S^1$ in the presence of a non-trivial holonomy. The diagrams of Figure \ref{fig:vac_pol_gauge_sector}  (for $N=2,3$)  add up to \cite{Chaichian:1996yr}
\begin{multline}
\Pi_{mn}^{{\mbox\scriptsize NA}\,ij}(p, \omega)=\delta^{ij}\frac{g^2 N}{2L}\sum_{q\in \mathbb Z}\int\frac{d^3 k}{(2\pi)^3}\\\times\frac{4\delta_{mn}P^2+2\left(P_m K_n+P_n K_m\right)+4K_m K_n-3P_m P_n-2(K+P)^2\delta_{mn}}{\left[k^2+\left(\frac{2\pi q+\mu}{L}\right)^2 \right]\left[(\pmb k+\pmb p)^2+\left(\frac{2\pi q+\mu}{L}+\omega\right)^2 \right]}\\
\\+ (\mu \rightarrow -\mu)\,.
\label{the nonabelian polarization}
\end{multline}
%

\subsection{The general form of the gluon propagator}
\label{The general form of the gluon propagator}
 In this subsection, we briefly review the procedure to sum an infinite number of bubble diagrams.  The Euclidean  bare gluon propagator  in the Feynman gauge 
on $\mathbb R^3 \times \mathbb S^1$ takes the form
\begin{eqnarray}
D^{ab\,,0}_{mn}=\frac{\delta^{ab}\delta_{mn}}{p^2+\omega^2}\,,
\end{eqnarray}
where $a,b$ are the color indices.  The vacuum polarization is defined as the difference between the inverse full gluon propagator and the inverse bare gluon propagator:
\begin{eqnarray}
\Pi^{ab}_{mn}=D^{-1\,ab\,,0}_{mn}-D^{-1\,ab}_{mn}\,.
\end{eqnarray}
Assuming that the polarization tensor is diagonal in the color indices, we can write the polarization tensor in the general form
\begin{eqnarray}
\Pi_{mn}^{ab}=\delta^{ab}\Pi_{mn}=\delta^{ab}\left(F{\cal P}_{mn}^L+G{\cal P}_{mn}^T\right)\,,
\label{general polarization}
\end{eqnarray}
where the projection operators ${\cal P}_{mn}^{\mbox{\scriptsize T}}$ and ${\cal P}_{mn}^{\mbox{\scriptsize L}}$ are defined as
\begin{eqnarray}
{\cal P}_{33}^T={\cal P}_{3\mu}^{\mbox{\scriptsize T}}=0\,,\quad {\cal P}_{\mu\nu}^T=\delta_{\mu\nu}-\frac{p_\mu p_\nu}{p^2}\,,\quad {\cal P}_{mn}^{\mbox{\scriptsize L}}=\delta_{mn}-\frac{P_m P_n}{p^2+\omega^2}-{\cal P}_{mn}^{\mbox{\scriptsize T}}\,.
\end{eqnarray} 
The projection operators obey the usual relations ${\cal P}^T{\cal P}^T={\cal P}^T$, ${\cal P}^L{\cal P}^L={\cal P}^L$, and ${\cal P}^T{\cal P}^L=0$. Next, one uses the fact that the polarization tensor is transverse, $P_m\Pi_{mn}=0$, to express $\Pi_{33}$ as
\begin{eqnarray}
\Pi_{33}=\frac{p_\mu p_\mu\Pi_{\mu\nu}}{p_3^2}\,.
\end{eqnarray}
Then, we set $m=n=3$ and $m=n=\mu$ in (\ref{general polarization}), to obtain
\begin{eqnarray}
F=\left(1+\frac{\omega^2}{p^2}\right)\Pi_{33}\,,\quad G=\frac{1}{2}\left(\Pi_{\mu\mu}-\frac{\omega^2}{p^2}\Pi_{33} \right)\,.
\label{F and G functions}
\end{eqnarray}
One can therefore express the full polarization tensor as a function of $\Pi_{33}$ and $\Pi_{\mu\mu}$. Notice that in order for the $F$ and $G$ functions to be non-singular, the $\Pi_{33}$ must vanish as $p^2$ for nonzero $\omega$. This will be one check of our results given in \eqref{final expression Pi QCD 00}. Finally, one can sum the polarization tensor to obtain the full propagator
\begin{eqnarray}
D_{mn}^{ab}=\delta^{ab}\left[\frac{1}{p^2+\omega^2-G}{\cal P}_{mn}^T+\frac{1}{p^2+\omega^2-F}{\cal P}_{mn}^L\right]\,.
\label{the resummed propagator with F and G}
\end{eqnarray}
As we will see in section \ref{The static limit, resummation and  absence of IR renormalons}, the full propagator on $\mathbb R^3 \times \mathbb S^1$ in the presence of a non-trivial holonomy does not suffer from any IR logarithmic singularity, and hence QCD(adj) on a compact circle is an IR renormalon free theory.

\section{Calculating the polarization tensor on $\mathbb R^3 \times \mathbb S^1$}
\label{Calculating the polarization tensor on a compact circle}

In this section, we compute the one-loop contribution from fermions and gauge bosons as given by (\ref{final expression Pi in cartan}) and  (\ref{the nonabelian polarization}). All the excited Kaluza-Klein modes cannot cause $IR$ problems as they have a Matsubara mass, and all infrared singularities, if any, will show up in the static limit $\omega=0$. Before taking the static limit, we compute the total polarization for QCD(adj) as a general function of $p$ and $\omega$. We postpone the discussion of the static limit to the next section.

\subsection{The one-loop fermion correction}

The fermions contribution to the vacuum polarization is given by  \eqref{final expression Pi in cartan} (see Figure \ref{fig:vac_polarization_ferm}). Keeping in mind that this expression is given for a single Dirac fermion, and ignoring the color indices, we find that the contribution from $n_W$ Weyl fermions is given by the expression
\begin{eqnarray}
\nonumber
\Pi^{W}_{mn}=\frac{n_W}{2}\Pi^{\mbox{\scriptsize D}}_{mn}\,.
\label{Pi expression for nW}
\end{eqnarray}
Before proceeding to the calculations of (\ref{final expression Pi in cartan}), we first examine the limit $L\rightarrow \infty$. Making the replacement 
\begin{eqnarray}
\frac{1}{L}\sum_{n \in \mathbb Z}\int\frac{d^3k}{(2\pi)^3}\rightarrow \int\frac{d^4K}{(2\pi)^4}, 
\end{eqnarray}
and performing the calculations using the dimensional-regularization method, by substituting $4\rightarrow D=4-\epsilon$, we obtain
\begin{eqnarray}
\Pi_{mn}^{L\rightarrow \infty}(P)=-\frac{2n_WNg^2}{3\left(4\pi\right)^2}(P^2\delta_{mn}-P_m P_n)\log P^2\,,
\label{standard polarization from fermions}
\end{eqnarray}
where we have ignored the $1/\epsilon$ piece accompanying the logarithm which can absorbed in a counter term. The expression (\ref{standard polarization from fermions}) is the standard background field textbook result in four dimensional field theory.

Now, we turn to the full calculation of (\ref{final expression Pi in cartan}). First, one needs to sum over the Kaluza-Klein modes. Such sums can be performed with the help of  the complex plane and the residue theorem.  In Appendix (\ref{The Matsubara sums}), we summarize and compute the list of the sums we encounter in this paper. Among the sums we denote by $S_0$, $S_1$, $S_2$, and $S_3$, only $S_0$ and $S_1$ are independent. In terms of $S_0$, $S_1$, and $S_2$ the $\Pi_{33}$ and $\Pi_{\mu\mu}$ components of the polarization tensor read:
\begin{eqnarray}
\nonumber
\Pi_{\mu\mu}^W&=&-2n_WNg^2 \int \frac{d^3k}{(2\pi)^3}\left[\left(-\pmb k\cdot \pmb p +2k^2\right)S_1 -3\omega S_2-3S_0 \right]\,,\\
\Pi_{33}^W&=&-2n_WNg^2 \int \frac{d^3k}{(2\pi)^3}\left[-\pmb k\cdot \pmb p S_1-2k^2S_1+S_2\omega+S_0\right]\,.
\end{eqnarray}
The  structure of the sums $S_1$ to $S_3$, which appear in  Appendix (\ref{The Matsubara sums}), takes the general form
\begin{eqnarray}
S=\mbox{Re}\left(\frac{1}{e^{pL-i\mu}-1}\right){\cal F}(pL, \omega L)+\frac{1}{2}{\cal F}(pL, -\omega L)\,,
\label{structure of the general sum text}
\end{eqnarray}
where ${\cal F}$ depends on the specific details of the sum. The first term in (\ref{structure of the general sum text}) is the contribution from the $\mu$-dependent part of the sum, while the second term is the vacuum part, $L\rightarrow \infty$. We see that the $\mu$-dependent  term can be obtained from the vacuum part upon replacing $\frac{1}{2} \rightarrow \mbox{Re}\left[\frac{1}{e^{pL-i\mu}-1}\right]$ and $\omega\rightarrow -\omega$ . This observation will prove to be  crucial for the cancellation of the IR divergences as we explain below.

Using the integrals in Appendix \ref{Appendix: Integrals}, we can express the polarization tensor in terms of the integrals $I_0$, $I_1$, $I_2$, and $I_3$:
\begin{eqnarray}
\nonumber
\Pi_{\mu\mu}^{W}&=&-\frac{2n_W N g^2}{L^2}\left\{-\left(1+\frac{2\omega L}{pL}\tan^{-1}\left(\frac{pL}{\omega L}\right)\right)I_0    +\left(\frac{p^2 L^2}{2}+\frac{\omega^2 L^2}{2}\right)I_1 \right.\\
&&\left.\quad\quad\quad+4I_2+2\omega L I_3    \right\}\,,
\end{eqnarray}
and
\begin{eqnarray}
\nonumber
\Pi_{33}^{W}&=&-\frac{2n_W N g^2}{L^2}\left\{-\left(1-\frac{2\omega L}{pL}\tan^{-1}\left(\frac{pL}{\omega L}\right)\right)I_0   +\left(\frac{p^2 L^2}{2}+\frac{\omega^2 L^2}{2}\right)I_1 \right.\\
&&\left.\quad\quad\quad-4I_2-2\omega L I_3    \right\}\,.
\end{eqnarray}
The integrals $I_0$ to $I_3$ result from integrating the sums $S_0$ to $S_3$ over $k$. Inheriting the sum structure, the IR behavior of these integrals can be studied by casting the integrals in the following form:\footnote{We stress that this form is appropriate only to study the IR behavior of the integrals and to show the cancellation of the logarithms between the vacuum and the volume dependent parts.  }
\begin{eqnarray}
I=I^V+\delta I=\int_0 dx\left[\frac{1}{2}+\mbox{Re}\left(\frac{1}{e^{x-i\mu}-1}\right)\right]{\cal G}(pL,\omega L,x)\,,
\label{general form of the integral}
\end{eqnarray}
where the function ${\cal G}(pL,\omega L,x)$ depends on the details of the integral. The first part of (\ref{general form of the integral}), ${\cal G}(pL,\omega L,x)/2$,  is the vacuum contribution to the integral, while  $\mbox{Re}\left(\frac{1}{e^{x-i\mu}-1}\right){\cal G}(pL,\omega L,x)$ is the $\mu$-dependent contribution. 

Now, we discuss a general feature of the integral (\ref{general form of the integral}) which is vital for the absence of the infrared renormalons.  The integral $\int dx {\cal G}(pL,\omega L,x)/2$ has an IR logarithmic divergence:
\begin{eqnarray}
I^V=\mbox{lim}_{q\rightarrow 0}\int_q dx \frac{{\cal G}(pL,\omega L,x)}{2}=\frac{{\cal N}}{2}\log P+...\,,
\label{vacuum part}
\end{eqnarray}
where  $P=\sqrt{p^2+\omega^2}$,  ${\cal N}$ is some number that depends on the explicit form of ${\cal G}$, and the dots represent terms that are not singular in the IR. On the other hand, the $\mu$-dependent part in (\ref{general form of the integral}) suffers from the same IR divergence which can be extracted by expanding $\mbox{Re}\left(\frac{1}{e^{x-i\mu}-1}\right)$ about $x=0$: $\mbox{Re}\left(\frac{1}{e^{x-i\mu}-1}\right)\cong-\frac{1}{2}+{\cal O}(x)$. Thus, we find
\begin{eqnarray}
\delta I&=&\mbox{lim}_{q\rightarrow 0}\int_q dx\mbox{Re}\left(\frac{1}{e^{x-i\mu}-1}\right){\cal G}(pL,\omega L,x)\\
&=&-\mbox{lim}_{q\rightarrow 0}\int_q dx \frac{{\cal G}(pL, \omega L,x)}{2}+...=-\frac{{\cal N}}{2}\log P+...\,.
\label{L dependent part}
\end{eqnarray}
Comparing (\ref{vacuum part}) with  (\ref{L dependent part}), we see that the IR parts cancel as we add $I^V$ to $\delta I$. This cancellation can also be seen immediately by computing the integrals $\delta I_0$ to $\delta I_3$.  To the best of our knowledge, these integrals are not known in the literature. In Appendix  \ref{eq:integrals_deriv}, we use a novel method to compute the integrals.  We list the final expressions of $I_0$ to $I_3$ in Appendix \ref{explicit form of the integrals}. From the explicit form of these integral, we find that the logarithms that come from the lower limit of the vacuum integrals $I^V$ cancel exactly with the corresponding logarithmic dependence of $\delta I$.

\subsection{The non-abelian part of QCD one-loop correction}

In this subsection, we repeat the same analysis we carried out above for the non-abelian case. At infinite circle radius we obtain from (\ref{the nonabelian polarization})
\begin{eqnarray}
\Pi_{mn}^{NA\,,L\rightarrow \infty}(P)=\frac{11g^2 N}{3(4\pi)^2}(P^2\delta_{mn}-P_m P_n)\log P^2\,,
\end{eqnarray}
thus we have the famous QCD $\beta$-function coefficient. 

Similar to the fermion case, the vacuum polarization of the non-abelian part (\ref{the nonabelian polarization}) can be expressed in terms of the sums $S_0$ to $S_3$ as:
\begin{eqnarray}
\nonumber
\Pi_{\mu\mu}^{NA}&=&\frac{g^2 N}{L}\int \frac{d^3 k}{(2\pi)^3}\left[\left(-8\pmb k\cdot \pmb p +3p^2+4k^2+6\omega^2\right)S_1 -6\left(S_0+2\omega S_2\right)\right]\,,\\
\Pi_{33}^{NA}&=&\frac{g^2 N}{L}\int \frac{d^3 k}{(2\pi)^3}\left(-\omega^2+2 p^2-4k^2-4\pmb k\cdot \pmb p \right)S_1+2S_0\,.
\end{eqnarray}
Next, we use the integrals in Appendix (\ref{Appendix: Integrals}) to find
\begin{eqnarray}
\nonumber
\Pi_{\mu\mu}^{NA}=\frac{g^2 N}{L^2}\left[-2\left(1+2\frac{\omega L}{pL}\tan^{-1}\left(\frac{pL}{\omega L}\right)\right)I_0+\left(7p^2L^2+10\omega^2L^2\right)I_1+8I_2+4\omega L I_3 \right]\,,\\
\end{eqnarray}
and
\begin{eqnarray}
\nonumber
\Pi_{33}^{NA}=\frac{g^2 N}{L^2}\left[-2\left(1-2\frac{\omega L}{pL}\tan^{-1}\left(\frac{pL}{\omega L}\right)\right)I_0+\left(4p^2L^2+\omega^2L^2\right)I_1-8I_2-4\omega L I_3 \right]\,.\\
\end{eqnarray}
Adding the contribution of fermions and the non-abelian part, we obtain the full polarization tensor $\Pi^{\mbox{\scriptsize QCD(adj)}}_{mn}=\Pi_{mn}^{NA}+\Pi_{mn}^{W}$ whose $\mu\mu$ and $33$ components read:
\begin{eqnarray}
\nonumber
\Pi^{\mbox{\scriptsize QCD(adj)}}_{\mu\mu}&=&\frac{g^2}{L^2} \left\{2N(n_W-1)\left(1+2\frac{\omega L}{pL}\tan^{-1}\left(\frac{pL}{\omega L}\right)\right)I_0\right.\\
\nonumber
&&\left.+N\left[(7-n_W)p^2L^2+(10-n_W)\omega^2L^2 \right]I_1+8N(1-n_W)I_2+4N(1-n_W)\omega LI_3\right\}\,,\\
\label{final expression Pi QCD mumu}
\end{eqnarray}
and
\begin{eqnarray}
\nonumber
\Pi_{33}^{\mbox{\scriptsize QCD(adj)}}&=&\frac{g^2}{L^2} \left\{2N(n_W-1)\left(1-2\frac{\omega L}{pL}\tan^{-1}\left(\frac{pL}{\omega L}\right)\right)I_0\right.\\
\nonumber
&&\left.+N\left[(4-n_W)p^2L^2+(1-n_W)\omega^2L^2 \right]I_1-8N(1-n_W)I_2-4N(1-n_W)\omega LI_3\right\}\,.\\
\label{final expression Pi QCD 00}
\end{eqnarray} 

Now, we can examine the functions $F$ and $G$ in (\ref{F and G functions}) as $pL\rightarrow 0$ for any non-zero value of the the Matsubara momentum $\omega$. We find that $F$ and $G$ do not suffer from any singularities as $pL\rightarrow 0$. In particular, the polarization tensor $\Pi_{33}^{\mbox{\scriptsize QCD(adj)}}/(pL)^2$ is a constant at $pL=0$.  In the next section, we closely examine the behavior of the polarization tensor as $pL\rightarrow 0$  at $\omega=0$. Due to the absence of any IR logarithmic singularities, we will find that the polarization tensor as well as the resummed propagator are well behaved in the IR.

\section{The static limit, resummation and  absence of IR renormalons}
\label{The static limit, resummation and  absence of IR renormalons}

In this section, we examine the polarization tensor and the resummed propagator when the external Matsubara frequency is set to zero. In fact, one expects the infrared problems to show up in this limit. To explain our point, let us first consider the case $L\rightarrow \infty$ where the polarization tensor behaves as  $\Pi_{mn}(P)=\frac{2g^2\beta_0}{(4\pi)^2}(P^2\delta_{mn}-P_mP_n )\log (P)$, where $\beta_0=\frac{N(11-2n_W)}{3}$. Now, Consider $n$ vacuum polarization graphs sandwiched between two external gluon at $\omega=0$. The full gluon propagator reads
\begin{eqnarray}
{\cal D}^{(n)}_{\mu\nu}(\omega=0,p)=\left[\left(\Pi_{33}\right)^n{\cal P}^0_{\mu\nu}+\left(\frac{1}{2}\Pi_{\mu\mu}\right)^n{\cal P}^T_{\mu\nu}\right]\frac{1}{\left(p^2\right)^{n+1}}\,,
\end{eqnarray}
and hence we obtain as we send $n \rightarrow \infty$
\begin{eqnarray}
\nonumber
{\cal D}_{mn}(p)&=&\frac{1}{p^2}\sum_{n=0}\left(\frac{2\beta_0 g^2}{(4\pi)^2}\log p\right)^k{\cal P}^L_{mn}+\frac{1}{p^2}\sum_{k=0}\left(\frac{2g^2\beta_0}{2(4\pi)^2}\log p\right)^k{\cal P}^T_{mn}\\
&=&\frac{{\cal P}^L_{mn}}{p^2\left[1-\frac{2g^2\beta_0}{(4\pi)^2}\log p\right]}+\frac{{\cal P}^T_{mn}}{p^2\left[1-\frac{2\beta_0 g^2}{(4\pi)^2}\log p\right]}\,.
\label{resummation at L infinity}
\end{eqnarray}
Therefore, we find that the above expression sufferers from IR logarithmic singularity as $p\rightarrow 0$ when setting $\omega=0$. It is this IR singularity that is responsible for the IR renormalons as we explained in the introduction. 

On the contrary, on $\mathbb R^3 \times \mathbb S^1 $ such IR logarithmic singularities disappear which in turn leaves the theory IR renormalon free. As we stressed in the previous section, the absence of logarithmic singularities is attributed to the cancellation of the IR logarithms between the vacuum and holonomy-dependent integrals. To further examine the situation, we explicitly write the polarization tensor of QCD(adj) on $\mathbb R^3 \times \mathbb S^1$ at $\omega=0$ in the limit $pL \ll 1$. From (\ref{final expression Pi QCD mumu}), (\ref{final expression Pi QCD 00}) and (\ref{approximation of integrals}) we obtain:
\begin{multline}
\Pi_{\mu\mu}^{\mbox{\scriptsize QCD(adj)}}(\omega L=0)=\frac{g^2 N p^2}{24\pi^2}(11-2n_W)\Bigg[2\log\left(\frac{L\Lambda_0}{4\pi}\right)\\-\left(\psi\left(\frac{\mu}{2\pi}\right)+\psi\left(1-\frac{\mu}{2\pi}\right) \right) \Bigg]-\frac{g^2Np^2}{36\pi^2}(n_W-1)\,,
\label{the magnetic mass}
\end{multline}
and
\begin{multline}
\Pi_{33}^{\mbox{\scriptsize QCD(adj)}}(\omega L=0)=g^2N\Bigg\{\frac{(n_W-1)}{6L^2}\left(-1+3\left(\frac{\mu}{\pi}-1\right)^2\right)\\
+\frac{11-2n_W}{48\pi^2}p^2\Bigg[2\log\left(\frac{\Lambda_0L}{4\pi}\right) \\-\left(\psi\left(\frac{\mu}{2\pi}\right)+\psi\left(1-\frac{\mu}{2\pi}\right) \right)\Bigg]\Bigg\}+\frac{g^2Np^2}{36\pi^2}(n_W-1) \,,
\label{the electric mass}
\end{multline}
where $\Lambda_0$ is some normalization scale that comes from the vacuum part of the integrals. 
Interestingly enough, we find $\Pi_{\mu\mu}^{\mbox{\scriptsize QCD(adj)}}(\omega L=0)\rightarrow 0$ as $pL \rightarrow 0$. In fact, recalling that this limit is equal to the gluon magnetic mass, we find that this result is not surprising since a photon mass term is forbidden thanks to the $U(1)^{N-1}$ gauge symmetry. As a bonus, one can use (\ref{the magnetic mass}) to obtain the running of the coupling constant of the $U(1)^{N-1}$ gauge theory due to the inclusion of Kaluza-Klein tower of excitations. Since the polarization term $\Pi_{\mu\nu}^{\mbox{\scriptsize QCD(adj)}}(\omega L=0)$ is just the wave function normalization for the background field ${\cal A}_\mu$ that  lives on $\mathbb R^3$ and is protected by the $U(1)^{N-1}$ gauge symmetry,  we find
\begin{multline}
\frac{1}{g^2_{g\,,\mbox{\scriptsize eff}}(L)}=\frac{1}{g^2_0}+\frac{N(11-2n_W)}{48\pi^2}\Bigg[\log\left(\frac{16\pi^2 }{L^2\Lambda_0^2}\right)\\+ \left(\psi\left(\frac{\mu}{2\pi}\right)+\psi\left(1-\frac{\mu}{2\pi}\right) \right)\Bigg]+\frac{N(n_W-1)}{72\pi^2}\,.
\label{running coupling}
\end{multline}
where $g_0$ is the bare coupling. Notice that the bare coupling term and the term which depends on the UV scale $\Lambda_{0}$ combine to give $1/g^2$ where $g^2$ is the usual one-loop coupling at scale $\sim L$. Setting $n_W=1$, we obtain the result (up to the renormalization scheme constants) of the running coupling in super Yang-Mills computed in \cite{Poppitz:2013zqa,Anber:2014lba} using the index theorem technology.

On the other hand, we have for the $33$ component of the vacuum polarization
\begin{eqnarray}
\Pi_{33}^{\mbox{\scriptsize QCD(adj)}}(\omega L=0,pL=0)=g^2N(n_W-1)\left[\frac{1}{6L^2}\left(-1+3\left(\frac{\mu}{\pi}-1\right)^2\right)\right]\,.
\label{Pi 33 at zero momentum and w}
\end{eqnarray} 
Since the gluon electric mass is defined as $m_g^2=-\Pi_{33}^{\mbox{\scriptsize QCD(adj)}}(\omega L=0,pL=0)$, we see that an electric mass is  generated by quantum corrections. The first term in (\ref{Pi 33 at zero momentum and w}) is a genuine electric mass term for the gluon. At the center-symmetric holonomy we have $\mu=\pi/N$, $N=2,3$, and therefor we find
\begin{eqnarray}
m_{sc}^2=\frac{g^2(1-n_W)}{3L^2}\left(N-6+\frac{6}{N}\right)\,.
\end{eqnarray}
For $n_W=1$ the electric mass  vanishes as expected for super Yang-Mills. In addition, we note that the second and third line in (\ref{the electric mass}) is the wave function normalization of the compact scalars, and it gives the scalar effective coupling
\begin{multline}
\frac{1}{g^2_{s\,,\mbox{\scriptsize eff}}(L)}=\frac{1}{g^2_0}+\frac{N(11-2n_W)}{48\pi^2}\Bigg[\log\left(\frac{16\pi^2 }{L^2\Lambda_0^2}\right)\\+ \left(\psi\left(\frac{\mu}{2\pi}\right)+\psi\left(1-\frac{\mu}{2\pi}\right) \right)\Bigg]-\frac{N(n_W-1)}{36\pi^2}\,.
\label{scalar running coupling}
\end{multline}
Notice that in the supersymmetric limit $n_W=1$ the coupling of the scalar is the same as the coupling of the 3D photon. This indeed has to be the case as they, upon photon dualization, combine to form the lowest component of the chiral multiplet.

Now, inserting the expressions (\ref{the electric mass}) and (\ref{the magnetic mass}) into the resummed propagator  (\ref{the resummed propagator with F and G}) we obtain
\begin{eqnarray}
\nonumber
{\cal D}_{mn}(\omega L=0)&=&\frac{{\cal P}^L_{mn}}{p^2\left[1-\frac{\beta_0g_0^2 N}{(4\pi)^2}\left[\log\left(\frac{\Lambda_0^2L^2}{4\pi^2}\right)-\left(\psi\left(\frac{\mu}{2\pi}\right)+\psi\left(1-\frac{\mu}{2\pi}\right) \right) \right]-N\frac{n_W-1}{36\pi^2} \right]+m_{sc}^2}\\
&+&\frac{{\cal P}^T_{mn}}{p^2\left[1-\frac{\beta_0g_0^2 N}{(4\pi)^2}\left[\log\left(\frac{\Lambda_0^2L^2}{4\pi^2}\right)-\left(\psi\left(\frac{\mu}{2\pi}\right)+\psi\left(1-\frac{\mu}{2\pi}\right) \right) \right]+N\frac{n_W-1}{72\pi^2} \right]}\,.
\end{eqnarray}
Comparing this expression to (\ref{resummation at L infinity}), we see that there is no IR logarithmic singularity, or any IR dependence on the external momentum\footnote{In other words the coupling stops running at scale $\sim L$} and hence QCD(adj) on $\mathbb R^3 \times \mathbb S^1$ is renormalon free.

\section{Conclusion}
\label{sec:conclusion}

In this work we have analyzed in detail the vacuum polarization of the massless photon on $\R^3\times \S^1$ in the center symmetric background. As we have shown in this theory all logarithmic dependence of the vacuum polarization as a function of the external momentum cancels and is cut off at short momentum scales, leading to the behavior very different than that of the theory on $\R^4$ where IR renormalons appear. The cancellation of logarithmic divergence we find is quite general, but it usually leads to IR problems which are worse than on $\R^4$. In the theory studied, however, all potentially IR dangerous effects are regulated by the IR cutoff scale $L$ -- the radius of the compact circle. 

This observation makes the recent connection of renormalons to non-perturbative saddles in \cite{Argyres:2012vv,Argyres:2012ka,Dunne:2012ae,Dunne:2012zk,Cherman:2013yfa,Cherman:2014ofa} more difficult to make. All hope is not lost, however, but the issue seems much more subtle, since the vanishing of the renormalon growth seems to be compensated by the corresponding diagram proliferation, as we discussed in the Introduction. Making this connection in Quantum Field Theory is very difficult as large orders of perturbation theory are not tractable. It may well be worth while to study the 2D, asymptotically free toy models such as $O(N)$, $CP(N-1)$ and the Principal Chiral Models and attempt to connect the renormalon growth to the diagram proliferation, as these theories reduce to quantum mechanics upon compactification, where a lot is known about the large orders of perturbation theory.

Although renormalon divergence no longer exists, we should emphasize that the position of the renormalon ambiguity on $\R^4$ in the Borel plane, which owes its existence to logarithms which no longer exist on $\R^3\times \S^1$, nevertheless had nothing to do with the bubble chain, but rather with the low momentum dependence of the $F$-function which resulted from the integration over the large fermion loop in the renormalon diagram (see \ref{sec:renormalons}). It is this function which dictated the structure of the OPE and the condensates which can appear in the theory, and it is very likely that this is still the case. So in order to understand how the condensates change as $L$ is changed, one may very well need to study the large fermion loop and the resulting $F$-function and its dependence on the momentum running into the chain.

Finally let us mention a curiosity about the large $N$ expansion. In the large $N$ limit it can be shown that only planar diagrams contribute in the perturbative expansion, and that they do not proliferate factorially but as a power \cite{Koplik:1977pf}. This is perfectly reasonable, as factorial growth of diagrams is associated with  instanton saddles, which are irrelevant in the large $N$ limit. However the theory on $\R^3\times \S^1$ with preserved center symmetry has additional saddles (i.e. instanton-monopoles) which carry $1/N$ of the instanton action, and are important in the large $N$ limit where $NL$ is kept fixed, i.e. the \emph{abelian large $N$ limit} (see Footnote \ref{fn:abelian limit}). The monopole--anti-monopole saddles will contribute to the ambiguity in the Borel plane which should then be canceled by the corresponding ambiguity in the perturbation theory. Since we have shown that this ambiguity does not come from the renormalon-type processes, it must come from the factorial growth of the number of diagrams. So the abelian large $N$ limit must have contributions from non-planar diagrams as well which will show factorial proliferation.

\acknowledgments

We would like to thank Mithat \"Unsal and Erich Poppitz for useful comments. The work of M.A. has been supported in part by a Discovery Grant from NSERC, Canada. The work of T.S. has been supported in part by BayEFG.  We would also like to thank the organizers of the workshop {\it Resurgence and Transseries in Quantum, Gauge and String Theories} which inspired many ideas.

\appendix

\section{Lie algebra and propagators on $\mathbb R^3 \times \mathbb S^1$ in the presence of holonomy}
\label{Appendix propagators}

In this appendix we review the Lie algebra, and then we derive the  propagators on $\mathbb R^3 \times \mathbb S^1$ in the presence of a non-trivial holonomy for a general gauge group. 

\subsection{Elements of the Lie algebra}

We consider a gauge group $G$ with Lie algebra $\left[T^a,T^b \right]=if^{abc}T^c$, where $a,b=1,2,...,\mbox{dim}(G)$.  The structure constants $f^{abc}$ are the Lie generators $\{T^a\}$ in the adjoint representation. Thus, they can be written as $\mbox{dim }(G)\times\mbox{dim}(G)$ matrices  $f^{abc}=-i(T_{\mbox{\scriptsize adj}}^b)_{ac}$. It will also prove convenient to use the Cartan-Weyl basis. In this basis one finds a maximal set of commuting generators  $\{H^i\}$ known as the Cartan generators:
\begin{eqnarray}
\left[H^i,H^j\right]=0\,,
\label{cartan generators commutator}
\end{eqnarray}
where $i,j=1,2,...,r$ and $r$ is the rank of the group. The remaining  $\mbox{dim}(G)-r$ generators are decomposed into and lowering $\{E_{-\pmb \beta}\}$ and raising $\{E_{\pmb \beta}\}$ generators which satisfy
\begin{eqnarray}
[ H^i, E_{\pm\pmb \beta}]=\pm  \beta^iE_{\pm\pmb \beta}\,.
\end{eqnarray}
The subscripts $\pm \pmb \beta$ denote the roots associated with the operators $E_{\pm \pmb \beta}$. A bold-face letter will be used to denote an $r$-component vector, so that $\pmb \beta=(\beta^1,\beta^2,...,\beta^r)$. Since $\{H^i\}$ form a  mutually commutating set, they can be represented by diagonal matrices. The non-zero structure constants are given in terms of the Cartan generators as
\begin{eqnarray}
f^{aib}=-i(T_{\mbox{\scriptsize adj}}^i)_{ab}=(H_{\mbox{\scriptsize adj}}^i)_{ab}= \beta^i\delta_{ab}\,,
\end{eqnarray}
where $a,b$ denote the remaining $\mbox{dim}(G)-r$ components. The generators $\{H^i\}$ and $\{E_{\pm \pmb \beta}\}$ are renormalized as
\begin{eqnarray}
\mbox{tr}_{\mbox{\scriptsize f}}\left[H^iH^j\right]=\frac{\delta^{ij}}{2}\,,\quad \mbox{tr}_{\mbox{\scriptsize f}}\left[E_{\pmb\beta}E_{\pmb\gamma }\right]=\frac{\delta_{\pmb \beta+\pmb \gamma=0}}{2}\,,
\end{eqnarray}
where f denotes the fundamental representation. Given this normalization,  we find 
\begin{eqnarray}
\mbox{tr}_{\mbox{\scriptsize adj}}\left[H^iH^j\right]=\delta^{ij}c_2\,,\quad \mbox{tr}_{\mbox{\scriptsize adj}}\left[f^{abc}f^{cda} \right]=\delta^{bd}c_2\,,
\end{eqnarray}
where $c_2$ is the dual Coxeter number. 

\subsection{The propagator in the presence of background holonomy}  

In this subsection, we derive the form of the propagator on $\mathbb R^3 \times \mathbb S^1$ for any gauge group $G$ in the presence of a non-trivial holonomy. This derivation works for scalars, fermions, and gauge fields. In the following, we derive the fermion propagator as an example. The Lagrangian of a Dirac fermion in the adjoint representation reads:
\begin{eqnarray}{\cal L}=\int d^4 x\bar \psi^a\gamma_m \left(\partial_m\delta^{ac}-i(T^{b}_{\mbox{\scriptsize adj}})_{ac}A^b_m\right)\psi^c\,.
\label{fermion propagator fabc}
\end{eqnarray}
A background holonomy can be introduced in terms of a constant field along the $S^1$ direction:
\begin{eqnarray}
A_m=\delta_{m3}A_3=\delta_{m3}A_3^bT_{\mbox{\scriptsize adj}}^b\,.
\end{eqnarray}
Then, it is trivial to see that the fermion propagator with periodic boundary conditions along the $S^1$ direction takes the form
\begin{eqnarray}
{\cal S}^{\mbox{\scriptsize F}\,n}_{A_3}=\frac{1}{\gamma_3\left(\frac{2\pi n}{L}+A_3^bT_{\mbox{\scriptsize adj}}^b\right)+\gamma_m l_m}\,.
\label{fermion prop first form}
\end{eqnarray} 

Equivalently, one can write the Lagrangian (\ref{fermion propagator fabc}) as
\begin{eqnarray}
{\cal L}=2\int d^4 x \mbox{tr}_{\mbox{\scriptsize f}}\left[ \bar \psi D_m\gamma_m \psi\right] \,,
\end{eqnarray}
where $D_m=\partial_m-i[A_m,\,\,]$. Then,  the spinor $\psi$ can be expanded in Cartan-Weyl basis as:
\begin{eqnarray}
\psi(x,x_0)=\frac{1}{L}\sum_{n \in \mathbb Z}e^{i\frac{2\pi n x^0}{L}}\left[\pmb \psi^n(x) \cdot \pmb H+\sum_{\pmb \beta_+}\psi_{\pmb \beta}^n(x) E_{\pmb \beta}+\sum_{\pmb \beta_+}\psi_{\pmb \beta}^{*n}(x) E_{-\pmb \beta}\right]\,,
\end{eqnarray}
where $\pmb H\equiv(H^1,H^2,...,H^r)$, $\{\pmb \beta_+\}$ is the set of the positive roots, and the generators $\{H^i\}$ and $\{E_{\pmb \beta}\}$ are in the fundamental representation. Next, a background holonomy can be introduced in terms of a constant field $\pmb \phi$ which lives along the Cartan generators in the $S^1$ direction:  
\begin{eqnarray}
A_m=A_3\delta_{m3}\equiv\frac{\pmb \phi \cdot \pmb H}{L}\delta_{m3}.
\end{eqnarray}
In general, this holonomy  breaks $G$ to $U(1)^r$, its maximal abelian subgroup. Using $[\pmb H, E_{\pm\pmb \beta}]=\pm \pmb \beta$, we obtain:
\begin{eqnarray}
\left[A_3,\psi\right]=\frac{1}{L}\sum_{n \in \mathbb Z}e^{i\frac{2\pi n x^0}{L}}\left[\sum_{\pmb\beta_+}\psi_{\pmb \beta}^n E_{\pmb \beta}\frac{\pmb\beta \cdot \pmb\phi}{L}-\sum_{\pmb\beta_+}\psi_{\pmb \beta}^{*n} E_{-\pmb \beta}\frac{\pmb\beta \cdot \pmb\phi}{L} \right]\,.
\end{eqnarray}
Then, we have
\begin{eqnarray}
\nonumber
&&\int_3^L dx^3 \mbox{tr}_f\left[ \bar \psi D_3\gamma_3 \psi\right]=\\
\nonumber
&&\frac{1}{L^2}\sum_{n,m \in \mathbb Z} \int_0^L d x^3 e^{-i \frac{2\pi (m-n)x^3}{L}}\left[\bar{\pmb \psi}^m(x) \cdot \pmb H+\sum_{\pmb \beta_+}\bar\psi_{\pmb \beta}^m(x) E_{\pmb \beta}+\sum_{\pmb \beta_+}\bar\psi_{\pmb \beta}^{*m}(x) E_{-\pmb \beta}\right]\\
\nonumber
&&\times i\gamma_3\left[ \frac{ 2\pi n}{L}\pmb \psi^n(x)\cdot \pmb H +\sum_{\pmb\beta'_+}\psi_{\pmb \beta'}^n(x) E_{\pmb \beta'}\left(-\frac{\pmb\beta' \cdot \pmb\phi}{L}+\frac{2\pi n}{L}\right)+\sum_{\pmb\beta'_+}\psi_{\pmb \beta'}^{*n}(x) E_{-\pmb \beta'}\left(\frac{\pmb\beta' \cdot \pmb\phi}{L}+\frac{2\pi n}{L}\right) \right]\,.\\
\end{eqnarray}
Using $\mbox{tr}_{\mbox{\scriptsize f}}[H^iH^j]=\delta_{\mu\nu}/2$, $\mbox{tr}_{\mbox{\scriptsize f}}\left[E_{\pmb \beta}E_{-\pmb \beta'} \right]=\delta_{\pmb\beta\pmb \beta'}/2$, we finally obtain 
\begin{eqnarray}
\nonumber
&&2\int d^4x \mbox{tr}_{\mbox{\scriptsize f}}\left[\bar \psi D_m\gamma_m \psi \right]=\\
\nonumber
&&\frac{1}{L}\int d^3 x\sum_{n \in \mathbb Z}\left\{
\sum_{\pmb\beta_+}\bar \psi^n_{\pmb\beta}\left[i\gamma_3\left(\frac{2\pi n}{L}+\frac{\pmb \beta \cdot \pmb \phi}{L}\right)+\gamma_m \partial_m \right]\psi^{*n}_{\pmb \beta}+\sum_{\pmb\beta_+}\bar \psi^{*n}_{\pmb\beta}\left[i\gamma_3\left(\frac{2\pi n}{L}-\frac{\pmb \beta \cdot \pmb \phi}{L}\right)+\gamma_m \partial_m \right]\psi^{n}_{\pmb \beta}\right.\\
&&\left.\quad\quad\quad\quad\quad+\bar{ \pmb\psi}^n\left(i\gamma_3\frac{2\pi n}{L}+\gamma_m\partial_m\right)\cdot\pmb\psi^n\right\}\,.
\label{final expression for fermions Lagrangian}
\end{eqnarray}
 The first and second terms in (\ref{final expression for fermions Lagrangian}) describe a tower of charged fermions under the broken $U(1)^r$ with masses $|\frac{2\pi n}{L}\pm\frac{\pmb \beta \cdot \pmb \phi}{L}|$. Then, the propagator of the charged fermions reads:
\begin{eqnarray}
{\cal S}^{\mbox{\scriptsize F}\,n}_{\pmb \beta}=\frac{1}{\gamma_3\left(\frac{2\pi n}{L}+\frac{\pmb \beta \cdot \pmb \phi}{L}\right)+\gamma_m l_m}\,.
\label{fermion prop second form}
\end{eqnarray}
The last term in (\ref{final expression for fermions Lagrangian})  describes  neutral fermions  under $U(1)^r$ with masses $\frac{2\pi n}{L}$. Since these particles do not couple to any of the $U(1)$ photons, they do not play a role in our analysis and we ignore them. Both form of the propagators (\ref{fermion prop first form}) and (\ref{fermion prop second form}) will be used in the present work. 

Similarly, one can obtain the gluon (in the Feynman gauge) and the ghost propagators in the Cartan-Weyl basis:
\begin{eqnarray}
D^{\mbox{\scriptsize GL}\,ij}_{mn}=\frac{\delta^{ij}\delta_{mn}}{k^2+\left(\frac{2\pi n}{L}+\frac{\pmb \beta \cdot \pmb \phi}{L}\right)^2}\,,\quad G^{\mbox{\scriptsize GH}\,ij}=\frac{\delta^{ij}}{k^2+\left(\frac{2\pi n}{L}+\frac{\pmb \beta \cdot \pmb \phi}{L}\right)^2}\,.
\end{eqnarray}
%
 
\section{The Matsubara sums}
\label{The Matsubara sums}

In the present work, we need to perform the sums 
\begin{eqnarray}
\nonumber
S_0(k,p, \omega;\mu)&=&\frac{1}{2L}\sum_{n \in Z}\frac{1}{(\pmb k+\pmb p)^2+\left(\frac{2\pi n+\mu}{L}+\omega\right)^2}+(\mu\rightarrow -\mu)\,,\\
\nonumber
S_1(k,p,\omega;\mu)&=&\frac{1}{2L}\sum_{n \in Z}\frac{1}{k^2+\left(\frac{2\pi n+\mu}{L}\right)^2}\frac{1}{(\pmb k+\pmb p)^2+\left(\frac{2\pi n+\mu}{L}+\omega\right)^2}+(\mu\rightarrow -\mu)\,,\\
\nonumber
S_2(k,p, \omega;\mu)&=&\frac{1}{2L}\sum_{n \in Z}\frac{\frac{2\pi n+\mu}{L}}{k^2+\left(\frac{2\pi n+\mu}{L}\right)^2}\frac{1}{(\pmb k+\pmb p)^2+\left(\frac{2\pi n+\mu}{L}+\omega\right)^2}+(\mu\rightarrow -\mu)\,,\\
\nonumber
S_3(k,p, \omega;\mu)&=&\frac{1}{2L}\sum_{n \in Z}\frac{\left(\frac{2\pi n+\mu}{L}\right)^2}{k^2+\left(\frac{2\pi n+\mu}{L}\right)^2}\frac{1}{(\pmb k+\pmb p)^2+\left(\frac{2\pi n+\mu}{L}+\omega\right)^2}+(\mu\rightarrow -\mu)\,.
\end{eqnarray}

The sum $S_1$ can be obtained by considering the engineered integral
\begin{eqnarray}
I=\frac{1}{2\pi i}\oint_C dz \frac{1}{e^{Lz-i\mu}-1}\frac{1}{k^2-z^2}\frac{1}{(\pmb k+\pmb p)^2-\left(z+i\omega\right)^2}\,,
\label{integral in complex plane}
\end{eqnarray}
where $C$ is a circle enclosing the complex plane at infinity. This integral vanishes since the integrand goes to zero at large values of $z$. Actually, (\ref{integral in complex plane}) can be computed using the residue theorem, which in turn can be used to obtain the result of the sum. The integrand has simple poles at
\begin{eqnarray}
 z_{1,2}=\pm k\,,\quad z_{3,4}=-i\omega\pm (\pmb k+\pmb p)\,,\quad z_{n+4}=i\frac{2\pi n+\mu}{L}\,. 
\end{eqnarray}
Calculating the residues and using the identity
\begin{eqnarray}
\frac{1}{e^{-x-i\mu}-1}=-1-\frac{1}{e^{x+i\mu}-1}\,,
\end{eqnarray}
we find
\begin{eqnarray}
\nonumber
S_1(k,p, \omega;\mu)&=&\frac{1}{2}\mbox{Re}\left[\frac{1}{e^{Lk-i\mu}-1}\frac{1}{k}\frac{1}{(\pmb k+\pmb p)^2-\left(k+i\omega\right)^2}\right.\\
\nonumber
&&\left.+\frac{1}{e^{L|\pmb k+\pmb p|-i\mu}-1}\frac{1}{|\pmb k+\pmb p|}\frac{1}{k^2-\left(|\pmb k+\pmb p|-i\omega\right)^2}\right]\\
\nonumber
&&+\frac{1}{2}\frac{1}{k}\frac{1}{|\pmb k+\pmb p|^2-\left(k-i\omega\right)^2}+\frac{1}{2}\frac{1}{| \pmb k+\pmb p|}\frac{1}{k^2-\left(|\pmb k+\pmb p|+i\omega\right)^2}\\
&&+(\mu\rightarrow -\mu)\,.
\label{the first sum}
\end{eqnarray}
 The  structure of this sum repeats in all sums, so we take a moment to comment on it. The first two lines in (\ref{the first sum}) is the contribution from the $\mu$-dependent part of the sum, thanks to the presence of the factor $\mbox{Re}\left[\frac{1}{e^{L\left|\vec l\right|-i\mu}-1}\right]$. While the last  line in (\ref{the first sum}) is the vacuum, $L\rightarrow \infty$, part.
Comparing the first two lines with the last one, we see that the $\mu$-dependent part can be obtained from the vacuum part upon replacing $\frac{1}{2} \rightarrow \mbox{Re}\left[\frac{1}{e^{L\left|\vec l\right|-i\mu}-1}\right] $ and $\omega\rightarrow -\omega$. This observation is crucial  for the cancellation of the logarithmic divergences, which in turn kills the infrared renormalons.

Using the same method we obtain for $S_0$
\begin{eqnarray}\label{the second sum}
S_0(k,p, \omega;\mu)&=&\frac{1}{2|\pmb k+\pmb p|}\left[\frac{1}{2}+\mbox{Re}\frac{1}{e^{|\pmb k+\pmb p| L-i\mu}-1} \right]+(\mu\rightarrow -\mu)\,.
\end{eqnarray}
$S_0$ and $S_1$ are the main sums one needs to perform. The rest of the sums can be obtained from $S_0$ and $S_1$ using simple algebra:
\begin{eqnarray}
\nonumber
S_2(k,p, \omega;\mu)&=&\frac{1}{2\omega}\left[S_0(k,0, \omega;\mu)-S_0(k,p, \omega;\mu)+(k^2- |\pmb k+\pmb p|^2-\omega^2) S_1(k,p, \omega;\mu) \right]\,,\\
S_3(k,p, \omega;\mu)&=&-k^2S_1(k,p, \omega;\mu)+S_0(k,p, \omega;\mu)\,.
\end{eqnarray}
%

\section{Integrals}
\label{Appendix: Integrals}

In this appendix, we list important integrals. First let us define $\delta I_0, \delta I_1,\delta I_2,\delta I_3$:
\begin{align}
\nonumber
\delta I_0(\mu)&=\frac{2}{(2\pi)^2}\int_0^\infty dx\; x\mbox{Re}\frac{1}{e^{x-i\mu}-1},,\\
\nonumber
\delta I_1(pL, \omega L;\mu)&=\frac{1}{(2\pi)^2}\frac{1}{2pL}\int_0^\infty  dx\mbox{Re}\frac{1}{e^{x-i\mu}-1}\log \left[ \frac{(2x+Lp)^2+\omega^2L^2}{(2x-Lp)^2+\omega^2L^2}\right]\,,\\
\nonumber
\delta I_2(pL, \omega L;\mu)&=\frac{1}{(2\pi)^2}\frac{1}{4pL}\int_0^\infty  dx\; x^2\;\mbox{Re}\frac{1}{e^{x-i\mu}-1}\log \left[ \frac{(2x+Lp)^2+\omega^2L^2}{(2x-Lp)^2+\omega^2L^2}\right]\,,\\
\nonumber
\delta I_3(pL, \omega L;\mu)&=\frac{1}{(2\pi)^2}\frac{1}{p L}\int_0^\infty  dx\; x\;\mbox{Re}\frac{1}{e^{x-i\mu}-1}\\&\hspace{4cm}\times\Bigg[\tan^{-1}\left(\frac{2x+p L}{\omega L}\right)
-\tan^{-1}\left(\frac{2x- p L}{\omega L}\right) \Bigg] \,.\label{app:eq:int_defs}
\end{align}
Using these definitions, we find  the integrals over the sums:
\begin{align}
\nonumber
&\int \frac{d^3 k}{(2\pi)^3}S_0=\frac{I_0}{L^2}\,,&&\int \frac{d^3 k}{(2\pi)^3} S_1=I_1\,,\\
\nonumber
&\int \frac{d^3 k}{(2\pi)^3}\pmb k \cdot \pmb p S_1=-\frac{p^2}{2}I_1\,,&&\int \frac{d^3 k}{(2\pi)^3}S_2=-\frac{\omega}{2}I_1\,,\\
&\int \frac{d^3 k}{(2\pi)^3}k^2 S_1=\frac{1}{L^2}\left(I_0+2I_2-\frac{1}{2}\omega^2 L^2 I_1+\omega L I_3-\frac{\omega L}{pL}\tan^{-1}\left(\frac{pL}{\omega L}\right)I_0\right)\,.
\span\omit\span\omit
\end{align}
%

\section{The results of the integrals}
\label{explicit form of the integrals}

In this appendix, we list the results of the integrals. We write the integrals from $I_0$ to $I_3$ as $I^V+\delta I$ to denote the vacuum and the $\mu$-dependent parts. The vacuum parts can be computed from the usual $\R^4$ methods, while the rest is given by
\begin{align}
\nonumber
&\delta I_{0}=\frac{-1+3\left(\frac{\mu}{\pi}-1\right)^2}{24}\;,\\
\nonumber
&\delta I_1=\frac{1}{2(2\pi)^2}\log\left(\frac{PL}{4\pi}\right)-\frac{1}{2(2\pi)^2}+\frac{1}{2(2\pi)^2}\frac{\omega L}{pL}\tan^{-1}\left(\frac{pL}{\omega L}\right)\\
\nonumber
&-\frac{1}{4\pi pL}\text{Im }\Bigg[\log\Gamma\left(\frac{\mu}{2\pi}+\frac{\mathcal PL}{4\pi}\right)+(\mu\rightarrow 2\pi-\mu)\Bigg]\,,\\
\nonumber
&\delta I_2=\frac{(pL)^2-3(\omega L)^2}{48 (2\pi)^2}\left(\log\left(\frac{PL}{4\pi}\right)-\frac{1}{3}\right)+\frac{3(pL)^2-(\omega L)^2}{48 (2\pi)^2}\;\frac{\omega L}{pL}\tan^{-1}\frac{p}{\omega}\nonumber\\ 
\nonumber
&\hspace{2.0cm}\frac{\delta I_0}{4}+\frac{\pi}{2 pL}\text{Im}\Bigg[\left(\frac{ \mathcal P L}{4\pi}\right)^2 \log\Gamma\left(\frac{\mu}{2\pi}+\frac{ \mathcal P L}{4\pi}\right)\nonumber\\
\nonumber
&\hspace{4.0cm}-\frac{2 \mathcal P}{4\pi}\psi_{-2}\left(\frac{\mu}{2\pi}+\frac{\mathcal  P L}{4\pi}\right)+2\psi_{-3}\left(\frac{\mu}{2\pi}+\frac{ \mathcal PL}{4\pi}\right)+(\mu\rightarrow 2\pi-\mu)\Bigg]\,,\\
\nonumber
&\delta I_3=\frac{\omega L}{4(2\pi)^2}\left(\log\frac{PL}{4\pi}-\frac{1}{2}\right)+\frac{(\omega L)^2-(pL)^2}{8(2\pi)^2 pL}\tan^{-1}\frac{pL}{\omega L}\nonumber\\&\hspace{0.8cm}+\frac{\text{sign }(\omega)}{8(2\pi)^2}\text{Im }\Bigg[-{4\pi  \mathcal P L}\log\Gamma\left(\frac{\mu}{2\pi}+\frac{ \mathcal P L}{4\pi}\right)+\left({4\pi}\right)^2\psi_{-2}\left(\frac{\mu}{2\pi}+\frac{ \mathcal P L}{4\pi}\right)\nonumber\\&\hspace{11cm}+(\mu\rightarrow 2\pi-\mu)\Bigg]\,.
\end{align}
where $\mathcal P=|\omega|+ip$.  

We also give approximations of the integrals $\delta I_{1,2,3}$ or $pL\ll 1$
\begin{align}
\nonumber
&\delta I_1{\approx}\frac{1}{2(2\pi)^2}\log\left(\frac{LP}{4\pi}\right)-\frac{1}{(2\pi)^2}\left(1-\frac{|\omega|}{p}\tan^{-1}\frac{p}{|\omega|}\right)\nonumber\\
\nonumber
&\hspace{3cm}-\frac{1}{(4\pi)^2}\mbox{Re}\Bigg[\psi\left(\frac{\mu}{2\pi}+\frac{(|\omega|+ip) L}{4\pi}\right)+(\mu\rightarrow 2\pi-\mu)\Bigg]\,,\\
\nonumber
&\delta I_2{\approx} \frac{(pL)^2-3(\omega L)^2}{48 (2\pi)^2}\left(\log\left(\frac{PL}{4\pi}\right)-\frac{1}{3}\right)+\frac{3(pL)^2-(\omega L)^2}{48 (2\pi)^2 }\frac{|\omega|}{p}\tan^{-1}\frac{p}{|\omega|}\nonumber\\
\nonumber
&\hspace{3cm}+\frac{1}{32(2\pi)^2}\text{Re }\Bigg[\left(\omega ^2+i \omega  p-\frac{p^2}{3}\right)L^2 \psi\left(\frac{\mu}{2\pi}+\frac{|\omega|L+i pL}{4\pi}\right)+(\mu\rightarrow 2\pi-\mu)\Bigg]+\frac{\delta I_0}{4}\,,\\
\nonumber
&\delta I_3{\approx}\frac{\omega L}{8(2\pi)^2}\left(2\log\frac{PL}{4\pi}-1\right)+\frac{(\omega L)^2-(pL)^2}{8(2\pi)^2p}\tan^{-1}\frac{p}{\omega}\nonumber\\&\hspace{3cm}-\frac{\text{sign }(\omega)}{8(2\pi)^2}\Re\Bigg[\Big(|\omega|+i p/2\Big)L\psi\left(\frac{\mu}{2\pi}+\frac{|\omega|L+i pL}{4\pi}\right)+(\mu\rightarrow 2\pi-\mu)\Bigg]\,.
\label{approximation of integrals}
\end{align}

\section{The computations of integrals}

\label{eq:integrals_deriv}

Here we derive the integrals $\delta I_{1,2,3}$. The integral $\delta I_1$ can be written as
\be
\delta I_1(q L,\omega L;a)=\frac{1}{2}\int_0^\infty \frac{dk}{(2\pi)^2} \frac{1}{q}\log\frac{(2k+q)^2+\omega^2}{(2k-q)^2+\omega^2}f_{\mu}(k)\,,
\ee
where $f_\mu(k)=\Re\frac{1}{e^{kL+i\mu}-1}$, we made explicit that the integral can depend on $qL$ and $\omega L$ only (which can be seen by substituting $x=kL$). 

To compute this integral we will differentiate $q \delta I_1$ with respect to $q$ and $L$, obtaining
\begin{multline}\label{eq:I0intermediate}
\partial_L\partial_q(2q\delta I_1)=-2\int_0^\infty \frac{dk}{(2\pi)^2} \frac{2k+q}{(2k+q)^2+\omega^2}\Re\frac{k}{4\sinh^2(k L+i \mu)}+(q\rightarrow -q)=\\=\int_{-\infty}^\infty dk\;. \frac{2k+q}{(2k+q)^2+\omega^2}\partial_L\frac{1}{2}\coth\left(\frac{kL+i \mu}{2}\right)+(q\rightarrow -q)\;.
\end{multline}
It is tempting to pull the $\partial_L$ in front of the integral, and close the contour from above, turning the integral into a sum using the residue theorem. However, we must be careful here, as the integral and the differential do not necessarily commute when the integral is infinite. To proceed, therefore, let us rewrite the above expression  as
\begin{multline}\label{eq:I0intermediate1}
\Bigg[\partial_{L}\int_{-\infty}^\infty dk \left(\frac{2k+q}{(2k+q)^2+\omega^2}-\frac{1}{2(k+i\mu/L)}\right) \frac{1}{2}\coth\left(\frac{kL+i\mu}{2}\right)\\+\int dk\;  \partial_L\left(\frac{L}{2(kL+i\mu )} \frac{1}{2}\coth\left(\frac{kL+i\mu}{2}\right)\right)+(q\rightarrow -q)\Bigg]\,.
\end{multline}
Notice that we pulled the $\partial_L$ out of the integral in the first line, as the integral on which it acts is convergent. In the second line, however, this trick is not allowed, as doing so would not yield the correct result. Indeed, pulling out the derivative might give a false impression that the integral is zero, as we can simply substitute $x=kL$ under the integral, which renders it independent of $L$. If we however notice that $\partial L=k\frac{\partial }{\partial (kL)}=k\partial_x$, treating $k$ as a constant, the integral in the second line reduces to
\be
 \frac{1}{L}\int dx\partial_x\left(\frac{x}{2(x+i\mu)}\frac{1}{2}\coth\left(\frac{x+i\mu}{2}\right)\right)=\frac{1}{2L}\;.
\ee
Closing the contour in the first line of \eqref{eq:I0intermediate1} from above, the expression \eqref{eq:I0intermediate1} becomes
\begin{multline}
\partial_L\partial_q(2q \delta I_1)=\frac{1}{(2\pi)^2}\partial_L\Bigg\{\sum_{n=1}^\infty\left[\frac{2(n-a)-ib}{(2(n-a)-ib)^2-c^2}-\frac{1}{2n}\right]\\+\frac{\pi}{4}\cot\left(\pi(a+ib/2+c/2)\right)+c.c.\Bigg\}+\frac{1}{L(2\pi)^2}\,,
\end{multline}
where we labeled
\be
a=\frac{\mu}{2\pi}\;,\qquad b=\frac{qL}{2\pi}\;,\qquad c=\frac{|\omega| L}{2\pi}\;.
\ee
The above sum is easily expressed in terms of the digamma function
\begin{multline}
\partial_L\partial_q(2q\delta I_1)=-\frac{1}{4(2\pi)^2}\partial_L\Bigg[\psi\left(1-a-i\frac{b}{2}-\frac{c}{2}\right)+\psi\left(1-a-i\frac{b}{2}+\frac{c}{2}\right)\\-\pi\cot\left(\pi\left(a+i\frac{b}{2}+\frac{c}{2}\right)\right)+c.c.\Bigg]+\frac{1}{(2\pi)^2L}\;,
\end{multline}
or, using the identity $\psi(1-x)-\pi\cot(\pi x)=\psi(x)$, we have
\begin{equation}
\partial_L\partial_q(2qI_1)=-\frac{1}{2(2\pi)^2}\partial_L\Re\Bigg[\psi\left(a+i\frac{qL}{4\pi}+\frac{|\omega| L}{4\pi}\right)+\psi\left(1-a-i\frac{qL}{4\pi}+\frac{|\omega| L}{4\pi}\right)\Bigg]+\frac{1}{(2\pi)^2L}\;,
\end{equation}
Integrating the above expression with respect to $q$ yields
\begin{equation}
\partial_L(2q\delta I_1)=-\frac{1}{2\pi}\partial_L\Im\Bigg[\frac{1}{L}\log\Gamma\left(a+i\frac{qL}{4\pi}+\frac{|\omega| L}{4\pi}\right)-\frac{1}{L}\log\Gamma\left(1-a-i\frac{qL}{4\pi}+\frac{|\omega| L}{4\pi}\right)\Bigg]+\frac{q}{(2\pi)^2L}\;,
\end{equation}
where we have set the integration constant (which depends on $\omega,L$) to zero, because we must have that $I_1$ is nonsingular as $q\rightarrow 0$, so that the RHS above needs to vanish in this limit. 
The above expression is finally integrated with respect to $L$ to yield
\begin{multline}
2\delta I_1=-\frac{1}{2\pi q L} \Im\Bigg[\log\Gamma\left(a+i\frac{qL}{4\pi}+\frac{|\omega| L}{4\pi}\right)\\-\log\Gamma\left(1-a-i\frac{qL}{4\pi}+\frac{|\omega| L}{4\pi}\right)\Bigg]+\frac{\log (LQ)}{(2\pi)^2}+C(\omega,q)\;,
\end{multline}
where $C(\omega,q)$ is the integration constant\footnote{Notice that we also added the constant $Q=\sqrt{q^2+\omega^2}$ under the logarithm in anticipation of the result.}, in general dependent on $\omega$ and $q$. In the limit $L\rightarrow \infty$, the above must vanish. Since the logaritham of the gamma function has an expansion
\be
\log(\Gamma(z))\approx z\log(z)-z
\ee
for large real part of $z$. Therefore, writing $|\omega|+i q=Qe^{i\phi}$, where $\phi=\tan^{-1}(q/|\omega|)$, we have that

\begin{multline}
\Im\Bigg[\log\Gamma\left(a+i\frac{qL}{4\pi}+\frac{|\omega| L}{4\pi}\right)-\log\Gamma\left(1-a-i\frac{q L}{4\pi}+\frac{|\omega| L}{4\pi}\right)\Bigg]\approx\\\approx 2\Im\left[\frac{Q L}{4\pi} e^{i\phi}\log\left(\frac{Q Le^{i\phi}}{4\pi}\right)-\frac{QL}{4\pi}e^{i\phi}\right]\\=\frac{qL}{2\pi}\log(QL/(4\pi))-\frac{qL}{2\pi}\left(1-\frac{|\omega|}{q}\tan^{-1}\frac{q}{|\omega|}\right)\,,
\end{multline}
so that
\be
C=-\frac{1}{(2\pi)^2}\log(4\pi)-\frac{1}{(2\pi)^2}\left(1-\frac{|\omega|}{q}\tan^{-1}\frac{q}{|\omega|}\right)
\ee
so that, finally
\begin{multline}
\delta I_1=-\frac{1}{4\pi qL}\Im\Bigg[\log\Gamma\left(a+i\frac{qL}{4\pi}+\frac{|\omega| L}{4\pi}\right)+(a\rightarrow 1-a)\Bigg]\\+\frac{\log \left(\frac{LQ}{4\pi}\right)}{2(2\pi)^2}-\frac{1}{2(2\pi)^2}\left(1-\frac{|\omega|}{q}\tan^{-1}\frac{q}{|\omega|}\right)\;.
\end{multline}

Since we are mostly interested in the limit $qL\ll 1$, before integrating with respect to $q$ we can assume that the $\psi$-function varies slowly with $q$ in this limit, and take it out of the integral. Then we obtain
\begin{multline}
\delta I_1\overset{qL\ll 1}{\approx}\frac{1}{2(2\pi)^2}\log\left(\frac{LQ}{4\pi}\right)-\frac{1}{(2\pi)^2}\left(1-\frac{|\omega|}{q}\tan^{-1}\frac{q}{|\omega|}\right)\\-\frac{1}{2(4\pi)^2}\Re\Bigg[\psi\left(a+\frac{(|\omega|+iq) L}{4\pi}\right)+(a\rightarrow 1-a)\Bigg]\,.
\end{multline}

Next, consider the integral
\be
\delta I_2/L^2=\frac{1}{4(2\pi)^2}\int dk \frac{k^2}{q}\log\frac{(2k+q)^2+\omega^2}{(2k-q)^2+\omega^2}f_a(k)\;.
\ee
Multiplying the above expression with $q$ and differentiating with respect to $q,L$, using  similar techniques one can show that it reduces to the expression
\begin{multline}
4(2\pi)^2\partial_L\partial_q (q\delta I_2/L^2)=\frac{1}{8}\partial_L \text{Re }\Bigg[(|\omega|+iq)^2\psi(a+\frac{|\omega| L+i q L}{4\pi})+(a\rightarrow 1-a)\Bigg]\\+(2\pi)^2\partial_L \delta I_0+\frac{q^2-\omega^2}{4L}\;\;.
\end{multline}
Integrating with respect to $L$ and demanding that the result vanishes for $L\rightarrow \infty$ 
\begin{multline}
4(2\pi)^2\partial_q (q\delta I_2/L^2)=\frac{1}{8 L^2}\text{Re }\Bigg[(|\omega|L+iqL)^2\psi\left(a+\frac{|\omega|L+i q L}{4\pi}\right)+(a\rightarrow 1-a)\Bigg]\\(2\pi)^2 \delta I_0+\frac{q^2-\omega^2}{4}\log\frac{QL}{4\pi}+\frac{\omega q}{2}\tan^{-1}\frac{q}{\omega}\,.
\end{multline}
Integrating over $q$, demanding that $q\delta I_2\rightarrow 0$ in this limit (i.e. that $\delta I_2$ is non-singular), we get 
\begin{multline}
4(2\pi)^2 \delta I_2/L^2=\frac{8\pi^3}{L^3 q}\text{Im }\Bigg[\left(\frac{|\omega| L+i q L}{4\pi}\right)^2\log\Gamma\left(a+\frac{|\omega| L+i qL}{4\pi}\right)\\-\frac{2(|\omega| L+i qL)}{4\pi}\psi_{-2}\left(a+\frac{|\omega| L+i q L}{4\pi}\right)\\+2\psi_{-3}\left(a+\frac{|\omega| L+ i qL}{4\pi}\right)+(a\rightarrow 1-a)\Bigg]\\+\frac{q^2-3\omega^2}{12}\left(\log\left(\frac{QL}{4\pi}\right)-\frac{1}{3}\right)+\frac{3q^2-\omega^2}{12 }\frac{|\omega|}{q}\tan^{-1}\frac{q}{|\omega|}+(2\pi)^2\delta I_0/L^2\,,
\end{multline}
where $\psi_{-2}$ and $\psi_{-3}$ are the polygamma functions of negative order, which can be defined by 
\be
\psi_{n-1}(x)=\int_0^x dx' \psi_{n}(x')\;, n\le 0\,.
\ee

For $qL\ll 1$, before integrating with respect to $q$, we can assume that the $\psi$-function varies slowly in this limit, so that we need only to integrate the function in front
\begin{multline}
\delta I_2/L^2\overset{qL\ll 1}{\approx} \frac{q^2-3\omega^2}{48 (2\pi)^2}\left(\log\left(\frac{QL}{4\pi}\right)-\frac{1}{3}\right)+\frac{3q^2-\omega^2}{48 (2\pi)^2 }\frac{|\omega|}{q}\tan^{-1}\frac{q}{|\omega|}+\frac{\delta I_0}{4L^2}\\+\frac{1}{32(2\pi)^2}\text{Re }\Bigg[\left(\omega ^2+i \omega  q-\frac{q^2}{3}\right) \psi\left(a+\frac{|\omega|L}{4\pi}\right)+(a\rightarrow 1-a)\Bigg]\,.
\end{multline}

Finally we compute
\be
\delta I_3/L=\int \frac{dk }{(2\pi)^2}\frac{f_a(k)k}{ q}\left(\tan^{-1} \frac{2k+q}{\omega}-\tan^{-1}\frac{2k-q}{\omega}\right)\;.
\ee
Again we can obtain a result by differentiating with respect to $q$ and $L$
\be
(2\pi)^2\partial_L\partial_q (qI_3/L)= \frac{\text{sign}(\omega)}{8}\partial_L\Bigg\{-(|\omega|+i q)\psi(a+\frac{|\omega| L+i q L}{4\pi})+(a\rightarrow 1-a)\Bigg]+\frac{\omega}{4L}\;.
\ee
By integrating with respect to $L$, and demanding that the result vanishes for $L\rightarrow \infty$, we get
\begin{multline}
(2\pi)^2\partial_q (qI_3/L)=\frac{\text{sign }(\omega)}{8}\Re\Bigg[-(|\omega|+i q)\psi\left(a+\frac{|\omega|+iq}{4\pi}\right)+(a\rightarrow 1-a)\Bigg]\\+\frac{|\omega|}{4}\log\frac{LQ}{4\pi}-\frac{q}{4}\tan^{-1}\frac{q}{|\omega|}\,.
\end{multline}
Integrating with respect to $q$, and demanding that $qI_3$ vanishes when $q\rightarrow 0$, we obtain 
\begin{multline}
\delta I_3/L=\frac{\text{sign }(\omega)}{8(2\pi)^2q}\text{Im }\Bigg[-\frac{4\pi}{L}(|\omega|+i q)\log\Gamma\left(a+\frac{|\omega| L+iqL}{4\pi}\right)\\+\left(\frac{4\pi}{L}\right)^2\psi_{-2}\left(a+\frac{|\omega| L+i qL}{4\pi}\right)+(a\rightarrow 1-a)\Bigg]\\
+\frac{\omega}{8(2\pi)^2}\left(2\log\frac{QL}{4\pi}-1\right)+\frac{\omega^2-q^2}{8(2\pi)^2q}\tan^{-1}\frac{q}{\omega}\,.
\end{multline}
For $qL\ll 1$ we can repeat the integration, assuming the $\psi$-function varies slowly and obtain
\begin{multline}
\delta I_3/L\overset{qL\ll 1}{\approx}-\frac{\text{sign }(\omega)}{8(2\pi)^2}\Re\Bigg[\Big(|\omega|+i q/2\Big)\psi\left(a+\frac{|\omega|L+i qL}{4\pi}\right)+(a\rightarrow 1-a)\Bigg]\\+\frac{\omega}{8(2\pi)^2}\left(2\log\frac{QL}{4\pi}-1\right)+\frac{\omega^2-q^2}{8(2\pi)^2q}\tan^{-1}\frac{q}{\omega}\,.
\end{multline}

\bibliography{QCDadj_renormalons}

\bibliographystyle{JHEP}

\end{document}